\documentclass[11pt]{article}

\RequirePackage{etex}

\usepackage[margin=1in]{geometry}   
\usepackage{amsfonts}
\usepackage{amsmath} 
\usepackage{amssymb}
\usepackage{centernot}
\usepackage{booktabs}
\usepackage{multirow}
\usepackage{algorithm}
\usepackage{algorithmicx}

\usepackage[onehalfspacing]{setspace} 

\usepackage{float}
\usepackage{caption}
\usepackage{graphicx}
\usepackage{adjustbox}
\usepackage{lscape}

\usepackage{adjustbox}

\usepackage{xcolor}

\usepackage{relsize}

\usepackage{bbm}
\usepackage{subfig}

\definecolor{forestgreen}{RGB}{34,139,34}
\usepackage{hyperref}
\hypersetup{
    colorlinks=true,
    linkcolor=magenta,
    filecolor=magenta, 
    citecolor=forestgreen,      
    urlcolor=blue
}

\usepackage{authblk}

\usepackage{amsthm}
\newtheorem{theorem}{Theorem}
\newtheorem{corrolary}{Corollary}

\usepackage{xpatch}
\makeatletter
\xpatchcmd{\proof}{\@addpunct{.}}{\@addpunct{:}}{}{}
\makeatother

\makeatletter
\def\@hangfrom#1{\setbox\@tempboxa\hbox{{#1}}%
      \hangindent 0pt
      \noindent\box\@tempboxa}
\makeatother

\makeatletter
\newcommand{\vast}{\bBigg@{3}}
\newcommand{\Vast}{\bBigg@{4}}
\makeatother

\makeatletter
\newcommand*{\indep}{%
  \mathbin{%
    \mathpalette{\@indep}{}%
  }%
}
\newcommand*{\nindep}{%
  \mathbin{
    \mathpalette{\@indep}{\not}
  }%
}
\newcommand*{\@indep}[2]{%
  \sbox0{$#1\perp\m@th$}
  \sbox2{$#1=$}
  \sbox4{$#1\vcenter{}$}
  \rlap{\copy0}
  \dimen@=\dimexpr\ht2-\ht4-.2pt\relax
  \kern\dimen@
  {#2}%
  \kern\dimen@
  \copy0 
} 
\makeatother

\DeclareMathOperator{\E}{\textnormal{\mbox{E}}}

\usepackage[utf8]{inputenc}
\DeclareUnicodeCharacter{200E}{}

\usepackage{tikz}
\usetikzlibrary{positioning,shapes.geometric}

\usepackage{setspace}
\usepackage{cite}

\usepackage[stable]{footmisc}

\usepackage{rotating}

\makeatletter
\def\@seccntformat#1{\@ifundefined{#1@cntformat}%
   {\csname the#1\endcsname\quad}  
   {\csname #1@cntformat\endcsname}
}
\let\oldappendix\appendix 
\renewcommand\appendix{%
    \oldappendix
    \newcommand{\section@cntformat}{\appendixname~\thesection\quad}
}
\makeatother

\usepackage[absolute,showboxes]{textpos}

\TPMargin{5pt}

\usepackage{thmtools, thm-restate}

\usepackage{datetime}

\usepackage{sectsty}
\subsectionfont{\noindent\textbf\itshape}
\subsubsectionfont{\normalfont\itshape}

\begin{document}


\title{\textbf{Systematically Missing Data in Causally Interpretable Meta-Analysis} \vspace*{0.3in} }

\author[1]{Jon A. Steingrimsson\footnote{Address for correspondence: email: \texttt{jon\_steingrimsson@brown.edu}.}}
\author[2]{David H. Barker}
\author[1]{Ruofan Bie}
\author[3-5]{Issa J. Dahabreh}

\affil[1]{Department of Biostatistics, School of Public Health, Brown University, Providence, RI}
\affil[2]{Department of Psychiatry, Rhode Island Hospital, Providence, RI}
\affil[3]{CAUSALab, Harvard T.H. Chan School of Public Health, Boston, MA}
\affil[4]{Department of Epidemiology, Harvard T.H. Chan School of Public Health, Boston, MA}
\affil[5]{Department of Biostatistics, Harvard T.H. Chan School of Public Health, Boston, MA}

\maketitle{}

\thispagestyle{empty}

\clearpage

\thispagestyle{empty}

\vspace*{1in}


\begin{abstract}
\noindent
\linespread{1.7}\selectfont Causally interpretable meta-analysis combines information from a collection of randomized controlled trials to estimate treatment effects in a target population in which experimentation may not be possible but covariate information can be collected from a simple random sample. In such analyses, a key practical challenge is systematically missing data when some baseline covariates are not collected in all trials. Here, we provide identification results for potential (counterfactual) outcome means and average treatment effects in the target population when covariate data are systematically missing from some of the trials in the meta-analysis. We propose three estimators for the average treatment effect in the target population, examine their asymptotic properties, and show that they have good finite-sample performance in simulation studies. We use the estimators to analyze data from two large lung cancer screening trials and target population data from the National Health and Nutrition Examination Survey (NHANES). To accommodate the complex survey design of the NHANES, we modify the methods to incorporate survey sampling weights and allow for clustering.

\vspace{0.3in}

\end{abstract}

\clearpage 
\section{Introduction}

When multiple randomized trials compare the same treatments, it is natural to want to learn about treatment effects by synthesizing evidence across trials. Meta-analysis is an umbrella term for quantitative methods for evidence synthesis \cite{borenstein2011introduction, schmid2020handbook}. When each trial included in a meta-analysis recruits participants from a different, typically ill-defined, underlying population and treatment effects are heterogeneous across populations, the ``summary'' estimates produced by standard meta-analysis methods do not have a clear causal interpretation because they cannot be interpreted as treatment effects in a well-defined target population \cite{dahabreh2020toward}. Building on methods for extending (generalizing or transporting \cite{dahabreh2019commentaryonweiss}) treatment effects from a single clinical trial to a target population of substantive interest \cite{cole2010, tipton2012, omuircheartaigh2014, pearl2014, dahabreh2020extending}, we recently proposed methods for ``causally interpretable meta-analysis'' that combines information from multiple trials to estimate treatment effects in a well-defined target population \cite{dahabreh2019efficient, dahabreh2020toward}.

These methods assume that a common set of covariates adequate to render the trials and the target population exchangeable is available from all the trials and the sample of the target population. In attempts to apply the methods in practice, we have found that this assumption is often not true \cite{barker2021causally} and instead one or more covariates are systematically missing, in the sense that they are available from some trials but not others. This situation is different from the usual within-trial missingness, where values are missing from some observations in a given trial but available from other observations in the same trial: with systematic missingness, for one or more covariates, values are missing  from all observations in a given trial. 

There is a large literature on methods for dealing with within-trial missing data \cite{rubin1976inference, molenberghs2014handbook, robins1994estimation}. Popular methods for handling within-trial missing data assume that there is a positive probability of observing the full data given any values of the observed data. If a covariate is not measured in a specific data source, the probability of observing the full data conditional on an observation being from that data source is zero, violating the assumptions needed for within-trial missing data methods. Thus, addressing systematically missing data requires different identifiability conditions and estimation procedures than those used for within-trial missing data. Previous work on systematically missing data has focused on conventional individual participant data meta-analysis methods and has only considered multiple imputation-based approaches for addressing missingness \cite{resche2013multiple, jolani2015imputation, kunkel2017comparison, jolani2018hierarchical, resche2018multiple}. The approaches allow all covariates collected in each trial to be used for analysis and therefore can account for variables that are collected differently across trials (e.g.,~different trials use different survey instruments or measurement devices). This has advantages over complete case analysis that either does not use trials with missing covariates (leading to loss in efficiency) or does not use covariates that are missing in some trials (leading to potential bias). Despite these advantages, however, existing methods have not been given a clear causal interpretation, are often only supported by intuitive arguments (e.g., without formal identification analysis), and their theoretical properties remain largely unknown. In fact, the vast majority of the imputation methods rely on chained equations where the final imputation model may not even correspond to a true joint density \cite{arnold1989compatible, arnold2002exact}.

Here, we propose methods for handling systematically missing data when extending inferences about potential (counterfactual) outcome means and average treatment effects from multiple trials to a target population that lacks outcome or treatment information. We provide identification results, propose g-formula, weighting, and augmented weighting (doubly robust) estimators for these causal estimands, and study large-sample behavior of the estimators. We also examine the finite-sample behavior of the estimators using simulations and implement them to analyze data from two trials of lung cancer screening to learn about the effects of the treatments on a nationally representative target population. The target population data are obtained from the National Health and Nutrition Examination Survey (NHANES) and we show how to modify our estimators to account for the NHANES complex survey design by incorporating the survey sampling weights and multi-stage clustering. 

\section{Data structure and causal estimands}
\label{sec:struct}

We have data from a collection of studies $\mathcal{S}$, indexed by $s = 1, \ldots, L$, evaluating the effect of the same treatments on the same outcome. Although we will refer to these data sources as ``trials'' in the remainder of the paper, we note here that the collection $\mathcal{S}$ can consist of observational studies or a mixture of trials and observational studies (provided the conditions stated below can be plausibly assumed to hold for them). Let $Y$ denote a univariate outcome (continuous, binary, or count) measured at the end of followup, $A$ the random variable for treatment assignment taking values in a finite set $\mathcal{A}$, $X$ the vector of baseline (pre-randomization and pre-treatment) covariates, and $S$ a random variable indicating which trial the data comes from. In the absence of any missing data, the data from trial $s \in \mathcal{S}$ is assumed to be $\{(X_i,A_i,Y_i,S_i=s): i = 1, \ldots, n_s\}$, where $n_s$ is the total number of observations in trial $s$. 

We also collect baseline covariate information from a separately obtained random sample of $n_0$ individuals from the target population of substantive interest: $\{X_i: i=1, \ldots, n_0\}$. We shall use $S = 0$ to indicate observations from the target population. We do not require treatment or outcome information from the sample of the target population -- this is a possible strength of our methods because they can be used when treatment and outcome data from the target population are not collected (e.g., due to cost or the need for specialized ascertainment procedures), are of insufficient quality (e.g., due to gross measurement error), or subject to confounding by unmeasured variables (e.g.,~because treatment is not randomly assigned in the target population). Let $n = \sum_{j=0}^L n_j$ be the total sample size for the composite dataset formed by appending data from the collection of trials and the sample of the target population. In the absence of missing data, the data available on participant $i \in \{1, \ldots, n\}$ would be a realization of
\begin{equation}
  \mathcal{F}_i =\begin{cases}
    (X_i, A_i, Y_i, S_i), & \text{if $S_i \in \mathcal{S}$}, \\
    (X_i,S_i = 0), & \text{if $S_i = 0$}.
  \end{cases}
\end{equation}
Hereafter, we refer to $\mathcal{F} = \{\mathcal{F}_i: i =1, \ldots, n\}$ as the full data. Let $Y_i^a$ denote the potential outcome for participant $i$ under intervention to set treatment to $a \in \mathcal{A}$ \cite{rubin1974, robins2000d}. We focus on estimating the potential outcome means in the target population $\E[Y^a|S=0]$, for $a \in \mathcal{A}$. Many causal estimands of interest are functions of potential outcome means; for example the average treatment effect comparing two treatments $a$ and $a^\prime$ in the target population is equal to the difference of the corresponding potential outcome means: $\E[ Y^a - Y^{a^\prime} | S = 0 ] = \E[Y^a|S=0] - \E[Y^{a^\prime}|S=0]$.

Suppose now that some covariates are not collected in some of the trials (i.e., one or more covariates are systematically missing from some trials in the collection $\mathcal{S}$) but all covariates are collected in the target population. Let $K$ denote the number of different missing data patterns of systematically missing covariates; $X^{(k)}$ the observed covariate vector under missingness pattern $k \in \{1, \ldots, K\}$; $X^{(-k)}$ the a vector of the the components of $X$ that are missing under missingness pattern $k \in \{1, \ldots, K\}$; and $\mathcal{S}^k$ be the set of trials with missingness pattern $k$. Note that we allow the systematic missingness to be potentially non-monotone across trials. The observed data on participant $i \in \{1, \ldots, n\}$ are a realization of
\begin{equation}
  \mathcal{O}_i =\begin{cases}
    (X_i^{(k)}, A_i, Y_i, S_i), & \text{if $S_i \in \mathcal{S}^k$}, \\
    (X_i,S_i = 0), & \text{if $S_i = 0$}.
  \end{cases}
\end{equation}
Define $n_k^* = \sum_{i=1}^n I(S_i \in \mathcal{S}^k)$ as the number of observations with missingness pattern $\mathcal{S}^k$ and assume that $\frac{n_k^*}{n} \longrightarrow p_k > 0$ as $n \longrightarrow \infty$. We use the notation $\Pr_k$ and $\E_k$ to denote probabilities and expectations, respectively, that are conditional on the data sources in $\mathcal{S}^k \cup \{0\}$. Throughout, we use $f(\cdot)$ to generically denote densities.

\section{Identification analysis}
\label{sec:Id}

\subsection{Identifiability conditions}

We will show that the potential outcome means in the target population are identifiable using the observable data under the following identifiability conditions:
\begin{enumerate}
\item[A1.] Consistency: If $A_i = a$, then $Y_i^a = Y_i$ for every individual $i$ in any of the trials or the target population.
\item[A2.] Within-trial exchangeability over treatment for trials with missingness pattern $k$: $Y^a \indep A|(X, S, S \in \mathcal{S}^k)$ for each $k \in \{1, \ldots, K\}$ and $a \in \mathcal{A}$.
\item[A3.] Positivity of treatment assignment for trials with missingness pattern $k$: For every treatment $a \in \mathcal{A}$ and all missingness patterns $k \in \{1, \ldots, K\}$, $\Pr_k[A=a|X^{(k)} = x^{(k)}, S \in \mathcal{S}^k] > 0$ for every $x^{(k)}$ such that $f(x^{(k)}, S \in \mathcal{S}^k) > 0$.
\item[A4.] Exchangeability over data source (transportability): $Y^a \indep S|X$ for each $a \in \mathcal{A}$. 
\item[A5.] Positivity of participation for trials for a given missingness pattern: $\Pr_k[S \in \mathcal{S}^k|X^{(k)} = x^{(k)}] > 0$ for each $k \in \{1,\ldots, K\}$ and for every $x^{(k)}$ that has a positive density in the target population, $f(x^{(k)}, S=0)>0$.
\item[A6.] $Y \indep X^{(-k)}|X^{(k)}, S \in \mathcal{S}^k, A=a$ for every treatment $a \in \mathcal{A}$ and each missingness pattern $k \in \{1, \ldots, K\}$.
\end{enumerate}
Condition A1 implies that there is no direct effect of participation in some specific trial on the outcome \cite{dahabreh2019generalizingb}. Conditions A2 and A3 are expected to hold when the treatment is randomized and are on the collection of trials with missingness pattern $k$. Conditions A4 and A5 are used for causally interpretable meta-analysis in the absence of systematically missing data \cite{dahabreh2019efficient} and condition A6 ensures that for each missingness pattern the observable data are sufficient to identify the potential outcome means in the target population. We note that positivity of trial participation (condition A5) is on the pooled set from all trials with missingness pattern $k$. This implies that we can draw inference about a target population that has a broader covariate distribution than any individual trial. For example, assume a missingness pattern has two trials and the only covariate needed for transportability to the target population is age. If the first trial enrolls people aged $50-75$ and the second trial enrolls people aged $40-70$, then the positivity condition is satisfied for a target population that has an age range of $40-75$. Similarly, the positivity of treatment assignment condition (A3) is weaker than assuming positivity of treatment assignment within each trial as it only requires at least one trial in $\mathcal{S}^k$ to have a positive probability of assigning participants with missingness pattern $k$ to treatment $a$. The conditions imply that for a fixed treatment $a \in \mathcal{A}$, the expectations $\E[Y|X^{(k)}, S=s, A=a]$ are equal for all trials $s \in \mathcal{S}^k$ \cite{dahabreh2019efficient}. Because the expectations $\E[Y|X^{(k)}, S=s, A=a]$ are identified from the observable data for each $s \in \mathcal{S}^k$, the equality of these expectations is testable. Throughout, to simplify exposition, we assume perfect adherence to the the assigned treatment strategy, no losses to followup, and no missing data. In practical applications, when these complications arise, our methods can be naturally combined with well-established approaches for addressing them.

\subsection{Identification}

In Supplementary Web Appendix \ref{Proof-ID}, we prove the following identification result: 
\begin{theorem}
\label{thm:ID}
Under Assumptions A1 through A6, for each missing data pattern $k \in \{1, \ldots, K\}$, the potential outcome mean in the target population, $\E[Y^a|S=0]$, can be written as the following observed data functional:
\begin{equation}
\label{Out-Id}
\psi_k(a) = \E_k[\E_k[Y|X^{(k)}, A=a, S \in \mathcal{S}^k]|S=0],
\end{equation}
or, equivalently, using the weighting representation
\begin{equation}
\psi_k(a) = \frac{1}{\Pr_k[S=0]} \E_k\left[\frac{\Pr_k[S=0|X^{(k)}] I(S \in \mathcal{S}^k, A= a)}{\Pr_k[A =a|X^{(k)}, S \in \mathcal{S}^k] \Pr_k[S \in \mathcal{S}^k|X^{(k)}]} Y \right] \label{IW-ID}.
\end{equation} 
\end{theorem}
From results \eqref{Out-Id} and \eqref{IW-ID}, we have the following corollary.
\begin{corrolary}
For all weight vectors $\{w_k: k = 1, \ldots, K\}$ satisfying $\sum_{k=1}^K w_k = 1$, the potential outcome mean in the target population can be identified by
\begin{equation*}
\sum_{k=1}^K w_k \psi_k(a)  = \sum_{k=1}^K w_k \E_k[\E_k[Y|X^{(k)}, A=a, S \in \mathcal{S}^k]|S=0],
\end{equation*}
or, equivalently, by the weighting expression
\begin{equation*}
\sum_{k=1}^K w_k \frac{1}{\Pr_k[S=0]} \E_k\left[\frac{\Pr_k[S=0|X^{(k)}] I(S \in \mathcal{S}^k, A= a)}{\Pr_k[A =a|X^{(k)},S \in \mathcal{S}^k] \Pr_k[S \in \mathcal{S}^k|X^{(k)}]} Y \right].
\end{equation*}
\end{corrolary}
We discuss the choice of the weights $\{w_1, \ldots, w_k\}$ in Section \ref{sec:est-inf}.

The sampling model for causally interpretable meta-analysis assumes a stratified sampling setup, where trial observations are treated as-if randomly sampled from a (typically unspecified) super-population stratified by $S$ with sampling fractions that are constant within a trial, can vary between trials, and are unknown to the investigators \cite{dahabreh2020toward}. Furthermore, we assume that the target population data are randomly sampled from the target population, separately from the trial samples and with unknown sampling fraction. When the sampling fractions vary and are unknown, the probabilities $\Pr_k[S \in \mathcal{S}^k|X^{(k)}]$ are not identifiable as the sample proportion of observations from $\mathcal{S}^k$ in the data does not in general reflect the probability of being from $\mathcal{S}^k$ in the super-population. Nevertheless, as we show in Supplementary Web Appendix \ref{app:sep}, the potential outcome mean is still identifiable. More specifically, we show that if $D$ is an indicator that is equal to one when an individual contributes data to the analyses and zero otherwise, the potential outcome mean can be written as
\begin{equation}
\label{Out-Id-2}
\psi_k(a) = \E_k[\E_k[Y|X^{(k)}, A=a, S \in \mathcal{S}^k, D = 1]|S=0, D=1],
\end{equation}
or using the weighting representation
\begin{equation}
\psi_k(a) = \frac{1}{\Pr_k[S=0|D=1]} \E_k\left[\frac{\Pr_k[S=0|X^{(k)},D=1] I(S \in \mathcal{S}^k, A= a)}{\Pr_k[A =a|X^{(k)},S \in \mathcal{S}^k, D=1] \Pr_k[S \in \mathcal{S}^k|X^{(k)}, D=1]} Y \Bigg | D=1\right] \label{IW-ID-2}.
\end{equation}  
Because all quantities in expressions \eqref{Out-Id-2} and \eqref{IW-ID-2} condition on data availability ($D=1$), the potential outcome means are still identifiable under the sampling model described above. It follows that the potential outcome means and treatments effects in the target population are identifiable under this sampling model and the expectations in Theorem \ref{thm:ID} can be interpreted as integrals with respect to the densities under the same sampling model.

\section{Estimation and inference}
\label{sec:est-inf}

\subsection{Estimation}
\label{sec:est}
Up to now, we have treated the weights $w_k$ as fixed. But they can be estimated using the data; we use $\widehat w_k$ to denote the potentially data-dependent weights. We denote the asymptotic limit of $\widehat w_k$ as $w^*_k>0$ for $k = 1, \ldots, K$ and we assume that, for each $k$, $\widehat w_k$ is $\sqrt{n}$-convergent, and $\sum_{k=1}^K \widehat w_k = 1$. By using plug-in estimators for the unknown quantities in expressions \eqref{Out-Id} and \eqref{IW-ID} we obtain two estimators for the potential outcome mean in the target population. The g-formula \cite{robins1986} (outcome model-based) estimator is
\begin{equation}
\label{G-Est}
\widehat \psi_{\text{\tiny g}}(a) = \sum_{k=1}^K \widehat w_k \frac{1}{n_0} \sum_{i=1}^n I(S_i = 0) \widehat g_{a,k}(X_i^{(k)}).
\end{equation}
Here, $\widehat g_{a,k}(X^{(k)})$ is an estimator for $\E_k[Y|X^{(k)}, A=a, S \in \mathcal{S}^k]$. The g-formula estimator $\widehat \psi_{{\tiny g}}(a)$ is consistent if all outcome models are correctly specified, that is, if $\widehat g_{a,k}(X^{(k)}) \overset{P}{\longrightarrow} \E_k[Y|X^{(k)}, A=a, S \in \mathcal{S}^k]$ for each $k = 1, \ldots, K$. 

The weighting estimator for the potential outcome mean in the target population is 
\begin{equation}
\label{IW-Est}
\widehat \psi_{\text{\tiny w}}(a) = \sum_{k=1}^K \widehat w_k \frac{1}{n_0} \sum_{i=1}^n \widehat o_a(X^{(k)}_i, A_i,S_i) Y_i,
\end{equation}
where, for each missingness pattern $k \in \{1, \ldots, K\}$, we define the weights $\widehat o_a(X^{(k)}, A,S)$ as \[
\widehat o_a(X^{(k)}, A,S) = \frac{(1- \widehat p_k(X^{(k)})) I(S \in \mathcal{S}^k, A= a)}{\widehat e_{a,k}(X^{(k)}) \widehat p_k(X^{(k)})}.
\]
Here, $\widehat p_k(X^{(k)})$ is an estimator for $\Pr_k[S \in \mathcal{S}^k|X^{(k)}]$ and $\widehat e_{a,k}(X^{(k)})$ is an estimator for $\Pr_k[A =a|X^{(k)}, S \in \mathcal{S}^k]$. The weighting estimator is consistent if both the models for $\widehat e_{a,k}(X^{(k)})$ and  $\widehat p_k(X^{(k)})$ are correctly specified, that is, if $\widehat e_{a,k}(X^{(k)}) \overset{P}{\longrightarrow} \Pr_k[A =a|X^{(k)}, S \in \mathcal{S}^k]$ and  $\widehat p_k(X^{(k)}) \overset{P}{\longrightarrow}\Pr_k[S \in \mathcal{S}^k|X^{(k)}]$, for each $k= 1, \ldots, K$.

The weighting estimator is not guaranteed to take values in the support of $Y$, but a ``normalized'' version \cite{robins2007comment, dahabreh2019generalizing} of the weighting estimator
\begin{equation*}
\widehat \psi_{\text{\tiny w,norm}}(a) = \sum_{k=1}^K \widehat w_k \frac{1}{\sum_{i=1}^n \widehat o_a(X^{(k)}_i, A_i,S_i)} \sum_{i=1}^n \widehat o_a(X^{(k)}_i, A_i,S_i) Y_i
\end{equation*}
is guaranteed to take values in the support of $Y$.


In Supplementary Web Appendix \ref{Inf-Func}, we derive the first-order influence function of $\psi(a)$ and using plug-in estimators into the unknown quantities results in the ``doubly robust'' \cite{bang2005doubly} estimator
\begin{equation}
\label{DR-Est}
\widehat \psi_{\text{\tiny DR}}(a) = \sum_{k=1}^K \widehat w_k \frac{1}{n_0} \sum_{i=1}^n \left( \widehat o_a(X^{(k)}_i, A_i,S_i) (Y_i - \widehat g_{a,k}(X^{(k)}_i)) + I(S_i = 0) \widehat g_{a,k}(X^{(k)}_i) \right).
\end{equation}
A normalized version of the doubly robust estimator is given by
\begin{equation*}
\widehat \psi_{\text{\tiny DR,norm}}(a) = \sum_{k=1}^K \widehat w_k \left(\frac{\sum_{i=1}^n \widehat o_a(X^{(k)}_i, A_i,S_i) (Y_i - \widehat g_{a,k}(X^{(k)}_i))}{\sum_{i=1}^n \widehat o_a(X^{(k)}_i, A_i,S_i)}  + \frac{1}{n_0} \sum_{i=1}^n I(S_i = 0) \widehat g_{a,k}(X^{(k)}_i) \right).
\end{equation*}
The normalized version of the doubly robust estimator is not guaranteed to take values in the support of $Y$, but will often have better finite sample performance when some weights are extreme \cite{hajek1971comment,robins2007}.
The g-formula and the weighting estimators are special cases of the doubly robust estimator obtained by setting $\widehat p_k(X^{(k)}) = 1$ and $\widehat g_{a,k}(X^{(k)}) = 0$, respectively. All three estimators $\widehat \psi_{{\tiny g}}(a), \widehat \psi_{\text{\tiny w}}(a),$ and $\widehat \psi_{\text{\tiny DR}}(a)$ can be calculated using the observed data as they do not require outcome or treatment information in the target population and for trials in the set of trials with missingness pattern $k$, that is $\mathcal{S}^k$, the estimators only depend on the observed covariate pattern $X^{(k)}$.

\subsection{Large-sample behavior of the doubly robust estimator}

We define some notation that will be useful in deriving the large-sample behavior of $\widehat \psi_{\text{\tiny DR}}(a)$. For each $k \in \{1, \ldots, K\}$ and general functions $g_{a,k}'(X^{(k)}), e_{a,k}'(X^{(k)}),\gamma_k',$ and $p'_k(X^{(k)})$, define 
\begin{align*}
L(g_{a,k}'(X^{(k)}), &e_{a,k}'(X^{(k)}),\gamma_k', p'_k(X^{(k)})) \\
&= \gamma_k' \left( \frac{(1- p'_k(X^{(k)})) I(S \in \mathcal{S}^k, A= a)}{e_{a,k}'(X^{(k)}) p'_k(X^{(k)})} (Y - g_{a,k}'(X^{(k)})) + I(S = 0) g_{a,k}'(X^{(k)}) \right).
\end{align*}
Define $\widehat \gamma_k = \frac{n_0+n_k}{n_0}$, which is a non-parametric estimator for $\gamma_{k} = \Pr_k[S=0]^{-1}$. For a random variable $W$ define
\[
\mathbb{P}_{k}(W) = \frac{1}{n_k+n_0}\sum_{i: S_i \in \{0\} \cup \mathcal{S}^{k}} W_i
\]
and $\mathbb{G}_{k}(W) = \sqrt{n}(\mathbb{P}_{k}(W) - \E_k[W])$.
Using these definitions, we can rewrite the estimator as
\[
\widehat \psi_{\text{\tiny DR}}(a) = \sum_{k=1}^K \widehat w_k \mathbb{P}_{k}\big(L(\widehat g_{a,k}(X^{(k)}), \widehat e_{a,k}(X^{(k)}),\widehat \gamma_k, \widehat p_k(X^{(k)}))\big).
\]
Let $g_{a,k}^*(X^{(k)}), e_{a,k}^*(X^{(k)}),\gamma_{k}, p^*_k(X^{(k)})$ be the asymptotic limits of $\widehat g_{a,k}(X^{(k)}), \widehat e_{a,k}(X^{(k)}),\widehat \gamma_k, \widehat p_k(X^{(k)})$, respectively.  We assume that all the limits $g_{a,k}^*(X^{(k)}), e_{a,k}^*(X^{(k)}),\gamma_{k}, p^*_k(X^{(k)})$ exist and 
\[
\mathbb{P}_{k}\big(L(\widehat g_{a,k}(X^{(k)}), \widehat e_{a,k}(X^{(k)}),\widehat \gamma_k, \widehat p_k(X^{(k)}))\big)
\]
is uniformly bounded. To derive the asymptotic distribution of $\widehat \psi_{\text{\tiny DR}}(a)$ we make the following assumptions:
\begin{enumerate}
\item[B1.] For each $k \in \{1, \ldots, K\}$, at least one of the two following conditions hold
\[
\widehat g_{a,k}(X^{(k)}) \overset{P}{\longrightarrow} \E_k[Y|X^{(k)}, S \in \mathcal{S}^k, A=a]
\]
or
\begin{align*}
\widehat p_k(X^{(k)}) &\overset{P}{\longrightarrow} {\Pr}_k[S \in \mathcal{S}^k|X^{(k)}]\\
\widehat e_{a,k}(X^{(k)}) &\overset{P}{\longrightarrow} {\Pr}_k[A =a|X^{(k)}, S \in \mathcal{S}^k]. 
\end{align*}
\item[B2.] For each $k \in \{1, \ldots, K\}$, the sequences $L\big(\widehat g_{a,k}(X^{(k)}), \widehat e_{a,k}(X^{(k)}),\widehat \gamma_k, \widehat p_k(X^{(k)})\big)$ and \\
$L\big(g_{a,k}^*(X^{(k)}), e_{a,k}^*(X^{(k)}),\gamma_{k}, p^*_k(X^{(k)})\big)$ are Donsker \cite{van1996weak}.
\item[B3.] 
\[
||L\big(\widehat g_{a,k}(X^{(k)}), \widehat e_{a,k}(X^{(k)}),\widehat \gamma_k, \widehat p_k(X^{(k)})\big) - L\big(g_{a,k}^*(X^{(k)}), e_{a,k}^*(X^{(k)}),\gamma_{k}, p^*_k(X^{(k)})\big)||_2 \overset{P}{\longrightarrow} 0.
\]
\item[B4.] $\E\big[L\big(g_{a,k}^*(X^{(k)}), e_{a,k}^*(X^{(k)}),\gamma_{k}, p^*_k(X^{(k)})\big)^2 \big] < \infty$.
\end{enumerate}
Assumption B1 reflects the model doubly robustness property as the estimator requires for each $k \in \{1, \ldots, K\}$ that at least one of the outcome model $\widehat g_{a,k}(X^{(k)})$ or both the model for trial participation $\widehat p_k(X^{(k)})$ and the treatment model $\widehat e_{a,k}(X^{(k)})$ are correctly specified. Note that different missingness patterns can satisfy different parts of assumption B1. In other words, the assumption can be satisfied if for some patterns we are only able to correctly specify the models for the outcome and for others we are only able to correctly specify the models for trial participation and treatment, provided that at least one of these groups of models is satisfied for each pattern. Assumption B2 follows from Donsker preservation theorems if all the individual estimators $\widehat g_{a,k}(X^{(k)}), \widehat e_{a,k}(X^{(k)}), \widehat p_k(X^{(k)})$ and their corresponding limits are Donsker, the models $\widehat e_{a,k}(X^{(k)})$ and $\widehat p_k(X^{(k)})$ are uniformly bounded away from zero, and $\widehat g_{a,k}(X^{(k)})$ is uniformly bounded. The Donsker assumption restricts the class of models, but includes many commonly used classes of models that are more flexible than standard parametric models \cite{horowitz2009semiparametric}.
\begin{theorem}
\label{DR-thm}
If assumptions B1 through B4 hold, then:
\begin{enumerate}
\item The doubly robust estimator is consistent, $\widehat \psi_{\text{\tiny \emph{DR}}}(a) \overset{P}{\longrightarrow} \psi(a).$
\item The doubly robust estimator has the asymptotic representation
\begin{equation}
\sqrt{n}\big(\widehat \psi_{\text{\tiny \emph{DR}}}(a) - \psi(a)\big) = \sum_{k=1}^K w^*_k\mathbb{G}_{k}\bigg(L(g_{a,k}^*(X^{(k)}), e_{a,k}^*(X^{(k)}),\gamma_{k}, p^*_k(X^{(k)}))\bigg) + REM + o_p(1), \label{As-Rep}
\end{equation}
where 
\begin{align*}
REM \leq &\sqrt{n} O_P\Bigg(\sum_{k=1}^K (\widehat w_k - w^*_k)+ \sum_{k=1}^K \Big|\Big| \widehat g_{a,k}(X^{(k)}) - \E_k[Y|X^{(k)}, S \in \mathcal{S}^k, A=a]\Big|\Big|_2  \\
&\times \Big( \Big|\Big|\widehat p_k(X^{(k)}) - {\Pr}_k[S \in \mathcal{S}^k|X^{(k)}] \Big|\Big|_2 + \Big|\Big|\widehat e_{a,k}(X^{(k)}) - {\Pr}_k[A =a|X^{(k)}, S \in \mathcal{S}^k]\Big|\Big|_2 \Big)\Bigg).
\end{align*}
\end{enumerate}
\end{theorem}
The second part of Theorem 2 gives the rate of convergence result
\begin{align*}
||\widehat \psi_{\text{\tiny DR}}(a) &- \psi(a)||_2 \leq O_P\Bigg(\frac{1}{\sqrt{n}} + \sum_{k=1}^K \Big|\Big| \widehat g_{a,k}(X^{(k)}) - \E_k[Y|X^{(k)}, S \in \mathcal{S}^k, A=a]\Big|\Big|_2  \\
&\times \Big( \Big|\Big|\widehat p_k(X^{(k)}) - {\Pr}_k[S \in \mathcal{S}^k|X^{(k)}] \Big|\Big|_2 + \Big|\Big|\widehat e_{a,k}(X^{(k)}) - {\Pr}_k[A =a|X^{(k)}, S \in \mathcal{S}^k]\Big|\Big|_2 \Big)\Bigg).
\end{align*}
This shows that the rate of convergence of the doubly robust estimator depends on how fast the rate of convergence of the nuisance estimators $\widehat g_{a,k}(X^{(k)}), \widehat p_k(X^{(k)}),$ and $\widehat e_{a,k}(X^{(k)})$ are. For example, if for each $k \in \{1, \ldots, K\}$ the combined rate of convergence of $\widehat g_{a,k}(X^{(k)})$ and $(\widehat p_k(X^{(k)}),\widehat e_{a,k}(X^{(k)}))$ is $\sqrt{n}$, then the rate of convergence for $\widehat \psi_{\text{\tiny DR}}(a)$ is $\sqrt{n}$. This allows the nuisance parameters to be estimated using procedures that converge at rates slower than $\sqrt{n}$, as long as the combined rate of convergence is $\sqrt{n}$ (rate double robustness \cite{rotnitzky2021characterization}). This is not the case for the outcome or weighting estimators that inherit the rate of convergence of their respective nuisance parameters, $\widehat g_{a,k}(X^{(k)})$, or $(\widehat p_k(X^{(k)}),\widehat e_{a,k}(X^{(k)}))$.

Although the Donsker assumption on the estimators for the nuisance parameters (Assumption B2) is more flexible than assuming a parametric model, it can be too restrictive for more data-adaptive estimators. In that case, sample splitting \cite{chernozhukov2018double, bickel1988estimating, robins2008higher} can be used to avoid any Donsker assumptions. The sample split doubly robust estimator is calculated by:
\begin{enumerate}
    \item Split the data into two mutually exclusive and exhaustive sets. Let  $\mathcal{B}_1$ and $\mathcal{B}_2$ be indexes from each set.
    \item For each missingness pattern, use observations with indexes in $\mathcal{B}_2$ to calculate the estimators $\widehat g_{a,k}^{(2)}(X^{(k)}), \widehat e_{a,k}^{(2)}(X^{(k)}),$ and $\widehat p^{(2)}_k(X^{(k)})$. Calculate 
    \begin{equation*}
\widehat \psi_{\text{\tiny DR}}^{(1)}(a) = \sum_{k=1}^K \widehat w_k \frac{1}{\sum_{i \in \mathcal{B}_1} I(S_i = 0)} \sum_{i \in \mathcal{B}_1} \left( \widehat o_a^{(2)}(X^{(k)}_i, A_i,S_i) (Y_i - \widehat g_{a,k}^{(2)}(X^{(k)}_i)) + I(S_i = 0) \widehat g_{a,k}^{(2)}(X^{(k)}_i) \right),
\end{equation*}
where $\widehat o_a^{(2)}(X^{(k)}, A,S)$ is defined by replacing $\widehat e_{a,k}(X^{(k)})$ and $\widehat p_k(X^{(k)})$ by $\widehat e_{a,k}^{(2)}(X^{(k)})$ and $\widehat p^{(2)}_k(X^{(k)})$ in the definition of $\widehat o_a(X^{(k)}, A,S)$.
    \item Calculate $\widehat \psi_{\text{\tiny DR}}^{(2)}(a)$ by repeating Step 2 with the roles of $\mathcal{B}_1$ and $\mathcal{B}_2$ switched.
    \item The sample split doubly robust estimator is defined as $\widehat \psi_{\text{\tiny DR}}^{(SS)}(a) = \frac{1}{2} \left(\widehat \psi_{\text{\tiny DR}}^{(1)}(a) + \widehat \psi_{\text{\tiny DR}}^{(2)}(a)\right)$.
\end{enumerate}
In Supplementary Web Appendix \eqref{sec:Don} we show that if assumptions B1, B3, and B4 hold (i.e., without requiring the Donsker assumption) the sample split doubly robust estimator is consistent and has the same asymptotic representation as in Theorem \ref{DR-thm}. Hence, the sample split doubly robust estimator is model and rate doubly robust without requiring any Donsker assumptions.

So far we have discussed the properties the weights $\{\widehat w_1, \ldots, \widehat w_k\}$ need to satisfy, without describing a procedure for choosing their values. In traditional individual participant data meta-analysis, the optimal choice of weights (defined in terms of minimum asymptotic variance) is to weight each study by the inverse of the study-specific variance estimator \cite{zeng2015random}. This suggests using weights that are proportional to the inverse of the estimated pattern-specific sampling variance for each term in the sum in equation \eqref{DR-Est} \cite{vo2019rethinking}. The optimality of inverse variance weights in meta-analysis, however, relies on independence between the study-specific estimators. In causally interpretable meta-analysis, trials with each missingness pattern are ``standardized'' to the same target population, which induces dependence between the estimators for different missingness patterns. As a result, inverse variance weighting will not be in general optimal. In Supplementary Web Appendix \ref{sec:opt-weights} we show that the weights that minimize the asymptotic variance of the doubly robust estimator involve solving an estimating equation that includes a term that depends on the asymptotic variance of the estimator for each missingness pattern and a term that represents the correlation induced by standardizing the estimators using the same sample from the target population.

\subsection{Inference}

If for each $k \in \{1, \ldots, K\}$, the estimators $\widehat g_{a,k}(X^{(k)})$, $\widehat p_k(X^{(k)})$, and $\widehat e_{a,k}(X^{(k)})$ converge at a fast enough rate such that $REM = o_p(1)$, then the estimator $\widehat \psi_{\text{\tiny DR}}(a)$ is asymptotically normally distributed. Confidence intervals for $\widehat \psi_{\text{\tiny DR}}(a)$ can be constructed using either sandwich variance estimators or using the nonparametric bootstrap.

\subsection{Comparing treatments that are not compared head-to-head in any trial}

The methods developed can easily be extended to compare target population efficacy of treatments that are not compared head-to-head in any trial (i.e., to perform ``indirect treatment comparisons'' as in ``network meta-analysis'' \cite{lumley2002network}). More specifically, we can restrict the estimation of each potential outcome mean to the set of trials that include the treatment under consideration. In conventional or pooled meta-analyses, it is difficult to contrast potential outcome means across studies because differences could result from differences in treatment efficacy or differences in the underlying study populations. On the contrary, the methods developed here allow for comparisons of treatments in the context of the same target population, provided the identifiability conditions hold.

\section{Simulations}
\label{sec:sim}

We conducted simulations to evaluate finite sample performance of the three estimators proposed in Section \ref{sec:est-inf}. We compared the g-formula estimator $\widehat \psi_{{\tiny g}}(a)$, the weighting estimator $\widehat \psi_{\text{\tiny w}}(a)$, and the doubly robust estimator $\widehat \psi_{\text{\tiny DR}}(a)$. For comparison, we also implemented a naive pooled estimator that simply averages the outcomes for those individuals assigned to treatment $a$ in the combined data from all trials. That is, the naive pooled estimator is calculated using the formula $\widehat \psi_{\text{\tiny pool}}(a) = \frac{\sum_{i=1}^n Y_i I(A_i =a, S_i \neq 0)}{\sum_{i=1}^n I(A_i =a, S_i \neq 0)}$. We also implemented ``complete-case'' versions of the g-formula, weighting, and the doubly robust estimators that only used data from the trials that collected information on all covariates (these estimators ignore all information from trials in which covariate data are systematically missing).

\subsection{Data generation}

We simulated the covariate vector from a five-dimensional mean-zero multivariate normal distribution with covariance matrix with element $(i,j)$ equal to $0.6^{|i-j|}$, for $i=1, \ldots, 5$, and $j=1, \ldots, 5$. Participation in any trial was simulated from a logistic regression model with 
\[
\Pr[S = 0|X] = 1 - \frac{\exp(1 + 0.2 X_1 + 0.2 X_2 + 0.2 X_3 + 0.1 X_1^2 + 0.1 X_2^2 + 0.1 X_3^2)}{1 + \exp(1 + 0.2 X_1 + 0.2 X_2 + 0.2 X_3 + 0.1 X_1^2 + 0.1 X_2^2 + 0.1 X_3^2)}.
\]
Observations with $S \neq 0$, were allocated to one of three trials using a multinomial logistic regression 
\[
S|(X,S \neq 0) \sim \text{Multinomial}\left((p_1,p_2, p_3), \sum_{i=1}^n I(S_i \neq 0)\right),
\]
where 
\begin{align*}
    p_1 &= \Pr[S=1|X,S\neq0] = \frac{\theta}{1+ \theta + \zeta} \\  
    p_2 &= \Pr[S=2|X,S\neq0] = \frac{\zeta}{1+ \theta + \zeta} \\  
    p_3 &= \Pr[S=3|X,S\neq0] = 1 - p_1 -p_2,
\end{align*}
with 
\begin{align*}
\theta &= \text{exp}\left(\log(1.3) X_1 + \log(1.3) X_2 + \log(1.3) X_3\right)\\
\zeta &=  \text{exp}\left(\log(0.8) X_1 +\log(0.8) X_2 +\log(0.8) X_3\right).
\end{align*}

For all three trials, treatment assignment was simulated from a Bernoulli distribution with parameter $0.5$, reflecting marginal randomization. 

We generated potential outcomes using
\begin{align*}
Y^{1} &= 1 + 0.2 X_1 + 0.2 X_2 + 0.1 X_1^2 + 0.1 X_2^2 + \varepsilon^1 \\
Y^{0} &= -0.2 X_1 - 0.2 X_4 I(S \neq 2) - 0.1 X_1^2 - 0.1 X_2^2 + \varepsilon^0,
\end{align*}
where $\varepsilon^a$ are independent $\mathcal{N}(0,1)$ for $a \in 
\{0,1\}$. The observed outcome was generated under consistency as $Y = Y^1 A + Y^0 (1-A)$. We considered two systematically missing data scenarios. In the first missing data scenario, all covariates were fully observed in trials $S=1$ and $S=2$, but $X_5$ was missing in trial $S = 3$. This results in two different missing data patterns ($K=2$). For the second missing data scenario, all covariates were observed in trial $S=1$, $X_4$ was missing in trial $S=2$, and $X_5$ was missing in trial $S=3$. This results in three different missing data patterns ($K=3$). The total sample size in the collection of trials and the target population was $n=2,000$ with an average size of the target population sample of $454$ and the average sample size in trials $1$, $2$, and $3$ of $540$, $540$, and $466$, respectively. We ran $1000$ simulations for each setting. The potential outcome means in the target population were $\E[Y^1|S=0] = 1.07$ and $\E[Y^0|S=0] = -0.094$ (calculated using a numerical approximation).

\subsection{Implementation and results}

For trial participation, we used (1) a logistic regression model that included linear and quadratic terms for all covariates that are collected in each trial and (2) a logistic regression model that only included linear terms. Because the model for participation in any trial is a logistic regression model that includes linear and quadratic terms, we expect that, for each missingness-pattern, a logistic regression model that includes both linear and quadratic terms can better approximate the correct model whereas the logistic regression model that includes only linear terms is more severely misspecified. For the outcome, we used a correctly specified linear regression model that included linear and quadratic main effects for all covariates that are collected in the trial and a misspecified model that only included linear terms. As the treatment assignment was marginally randomized, any model that includes an intercept is correctly specified. In the simulations we used a main effects logistic regression estimator to estimate $\Pr_k[A =a|X^{(k)}, S \in \mathcal{S}^k]$. Estimating the treatment assignment can improve precision in randomized controlled trials compared to using the true value \cite{steingrimsson2017improving}. 

In the main part of the manuscript, to combine estimates across missing data patterns, we use weights that are proportional to the total sample size of the trials in a given missing data pattern $\mathcal{S}^k$. Additional simulations show that selecting the weights proportional to the sample size, proportional to the inverse of the estimated variance for each term in equation \eqref{DR-Est}, or using the optimal weights in terms of minimizing the asymptotic variance resulted in almost identical performance in terms of both bias and standard deviation (results not shown).

Table \ref{main-table} shows results from $1,000$ simulations where the estimators are compared in terms of bias and standard deviation. The naive pooled estimator was biased for the potential outcome mean in the target population for both missing data scenarios and both treatments. The estimators $\widehat \psi_{{\tiny g}}(a)$, $\widehat \psi_{\text{\tiny w}}(a)$, and $\widehat \psi_{\text{\tiny DR}}(a)$ were approximately unbiased when the outcome model was correctly specified and the model for trial participation included both linear and quadratic terms. When the outcome model was misspecified, the g-formula estimator was biased and when the probability of trial participation included only first order terms the weighting estimator was biased. The doubly robust estimator was unbiased if either the outcome model was correctly specified or the model for trial participation included both linear and quadratic terms. The g-formula estimator had the smallest variance, followed by the doubly robust estimator, and the weighting estimator had the largest variance. Using only data from trials with complete covariate information resulted in loss in efficiency for all three estimators compared to using the estimators described in Section \ref{sec:est}. 

\begin{table}[]
\begin{tabular}{|l|c|c|c|c|l|l|l|l|}
\hline
\multirow{2}{*}{}{$K$} &  
\multirow{2}{*}{}{Estimator} & \multirow{2}{*}{}{\shortstack[c]{ Correct\\ Outcome model}} &
\multirow{2}{*}{}{\shortstack[c]{Quadratic Trial \\ Participation Model}} &
\multirow{2}{*}{}{\shortstack[c]{Restricted to \\ Complete Trials}} &
\multicolumn{2}{c}{$a=0$} &
\multicolumn{2}{c|}{$a=1$} \\ \cline{6-9}
& & & & & Bias & SD & Bias & SD \\  \hline
\multirow{12}{*}{2} & Naive &   &   &   & -5.81  & 3.04 & 7.60   & 3.18 \\ \cline{2-9}
                      & GF    & x &   &   & -0.44  & 1.83 & 0.094  & 1.97 \\ \cline{2-9}
                      & GF    &   &   &   & -1.74  & 1.92 & 1.39   & 1.97 \\ \cline{2-9}
                      & GF    & x &   & x & -0.67  & 2.19 & 0.13   & 2.33 \\ \cline{2-9}
                      & W     &   & x &   & -0.54  & 1.97 & 0.089  & 1.97 \\ \cline{2-9}
                      & W     &   &   &   & -2.64  & 2.01 & 2.32   & 1.96 \\ \cline{2-9}
                      & W     &   & x & x & -0.67  & 2.32 & 0.089  & 2.28 \\ \cline{2-9}
                      & DR    & x & x &   & -0.44  & 1.88 & 0.085  & 1.97 \\ \cline{2-9}
                      & DR    &   & x &   & -0.45  & 1.92 & -0.098 & 1.96 \\ \cline{2-9}
                      & DR    & x &   &   & -0.45  & 1.88 & 0.098  & 1.97 \\ \cline{2-9}
                      & DR    &   &   &   & -3.00  & 1.96 & 2.60   & 1.93 \\ \cline{2-9}
                      & DR    & x & x & x & -0.67  & 2.24 & 0.12   & 2.28 \\ \hline
\multirow{12}{*}{3} & Naive &   &   &   & -5.81  & 3.04 & 7.60   & 3.18 \\ \cline{2-9}
                      & GF    & x &   &   & -0.45  & 1.88 & 0.11   & 1.98 \\ \cline{2-9}
                      & GF    &   &   &   & -1.74  & 1.92 & 1.39   & 1.97 \\ \cline{2-9}
                      & GF    & x &   & x & -0.020 & 3.13 & 0.094  & 3.04 \\ \cline{2-9}
                      & W     &   & x &   & -0.58  & 2.01 & 0.094  & 1.96 \\ \cline{2-9}
                      & W     &   &   &   & -2.68  & 2.06 & 2.37   & 1.97 \\ \cline{2-9}
                      & W     &   & x & x & -0.058 & 3.35 & 0.098  & 3.13 \\ \cline{2-9}
                      & DR    & x & x &   & -0.45  & 1.88 & 0.098  & 1.96 \\ \cline{2-9}
                      & DR    &   & x &   & -0.44  & 1.92 & 0.11   & 1.98 \\ \cline{2-9}
                      & DR    & x &   &   & -0.45  & 1.87 & 0.11   & 1.98 \\ \cline{2-9}
                      & DR    &   &   &   & -3.13  & 1.97 & 2.73   & 1.95 \\ \cline{2-9}
                      & DR    & x & x & x & -0.020 & 3.18 & 0.089  & 3.09 \\ \hline
\end{tabular}
\caption{Comparisons of the naive pooled $\widehat \psi_{\text{\tiny pool}}(a)$ (Naive), g-formula $\widehat \psi_{{\tiny g}}(a)$ (GF), weighting $\widehat \psi_{\text{\tiny w}}(a)$ (W), and doubly robust $\widehat \psi_{\text{\tiny DR}}(a)$ (DR) estimators in terms of bias and standard deviation (SD). To facilitate numerical comparisons, the bias and standard deviation estimators are multiplied by $\sqrt{n} = 44.7$. Restricted to complete trials refers to using only trials with no systematic missing data in the analysis. Correct outcome model refers to using a correctly specified outcome model and quadratic trial participation model refers to using a model for trial participation that includes linear and quadratic terms. $K$ refers to the the number of missingness-patterns.}
\label{main-table}
\end{table}

\section{Meta-analysis of lung cancer screening trials}
\label{sec:DA}

To illustrate the application of the methods, we use data from two lung cancer screening trials to estimate a potential outcome mean associated with chest radiography screening in a nationally representative target population of individuals who would be eligible for lung cancer screening. 

\subsection{Description of trial and target data}

The National Lung Screening trial (NLST) enrolled people aged 55 to 74 that had $\geq$ 30 pack-year history who were current smokers or had quit within the past 15 years. Participants were randomized to screening with computed tomography or chest radiography \cite{national2011national,national2010baseline}. The trial showed substantial benefits of lung cancer screening and its results have informed national lung cancer screening guidelines \cite{moyer2014screening, krist2021screening}. The Prostate, Lung, Colorectal, and Ovarian (PLCO) Cancer Screening Trial enrolled participants aged 55-74 with no prior history of prostate, lung, colorectal, or ovarian cancer \cite{oken2011screening, oken2005baseline}. The trial randomized participants to either a control arm that received usual care or a treatment arm where participants received screening for several cancers including screening for lung cancer using chest radiography. We used the binary outcome of whether a participant was diagnosed with lung cancer within six years from study entry.

The target population data came from the 2003-2004 NHANES, a cross-sectional survey with a multi-stage clustering design evaluating the health and nutritional status of non-institutionalized US adults and children \cite{johnson2013national}. NHANES oversampled certain subgroups (including both racial and ethnic subgroups), resulting in data that require the use of sampling weights to represent the underlying target population \cite{curtin2012national}. In the next subsection we describe how the g-formula, weighting, and doubly robust estimators can be modified to handle a weighted sampling design in the target population.

The target population we focused on is people recommended for lung cancer screening in US. As the NLST eligibility criteria were very similar to the population recommended for screening by the U.S. Preventive Services Task Force guidelines \cite{moyer2014screening, krist2021screening}, we restricted the NHANES data to only include individuals who met the NLST eligibility criteria (and have data on smoking habits that allow us to verify the eligibility criteria). Table \ref{tab:DA-app} in Supplementary Web Appendix \ref{app:da} shows summary statistics of the covariates used in the analysis stratified by data-source. The table shows that NHANES participants are on average less educated, have more comorbidities than participants in the two trials, and are less likely to live with a smoker compared to NLST participants. Because both trials evaluated chest radiography screening, we focused on estimating the potential outcome mean of that intervention in the NHANES target population (because screening with computerized tomography and usual care were each evaluated in only one of the trials, previously described transportability methods \cite{dahabreh2020extending} would suffice to transport potential outcome means under these interventions). 

Information on whether a participant lived with a smoker and if the participant had a history of asthma was collected in both the NLST and NHANES data but not in the PLCO data, resulting in systematically missing data; all three data sources collected the other covariate information listed in Table \ref{tab:DA-app} in Supplementary Web Appendix \ref{app:da}. Furthermore, both NLST and NHANES collected average number of cigarettes smoked per day as a continuous variable while in PLCO it was collected as a categorical variable. Thus, the transportability analysis from NLST to NHANES used average number of cigarettes smoked per day as a continuous variable while the transportability analysis from PLCO to NHANES used it as a categorical variable. For simplicity, we restrict the analysis to observations that have no within trial or target missing covariate data and trial participants that were not censored in the first six years. 
This resulted in $22,841$ observations from NLST, $17,639$ observations from PLCO, and $219$ observations from NHANES (representing roughly $8.5$ million participants) being used in our analysis. To implement the estimators, we used main effects logistic regression for the model for trial participation, the treatment assignment, and the outcome model, where the model for trial participation is weighted by the sampling weights as described in the next section.

\subsection{Accounting for survey sampling weights and multi-cluster sampling in the target population}
\label{app:weights}
Let $\eta_i$ denote the survey sampling weight associated with observation $i$ in the target population. We set these weights to one for all observations in the collection of trials. To account for the sampling weights in the data from the target population we modify the g-formula estimator:
\begin{equation}
\label{G-Est}
\widehat \psi^\eta_{{\tiny g}}(a) = \sum_{k=1}^K  \widehat w_k \frac{1}{\sum_{i=1}^n \eta_i I(S_i = 0)} \sum_{i=1}^n \eta_i I(S_i = 0) \widehat g_{a,k}(X_i^{(k)}).
\end{equation}
Because $\widehat g_{a,k}(X^{(k)})$ is an estimator for $\E_k[Y|X^{(k)}, A=a, S \in \mathcal{S}^k]$, it is estimated using only trial data; therefore, the procedure for estimating $\widehat g_{a,k}(X^{(k)})$ does not need to be modified to account for the sampling weights.

Furthermore, we also modify the weighting estimator to account for the sampling weights:
\begin{equation}
\label{IW-Est}
\widehat \psi^\eta_{\text{\tiny w}}(a) = \sum_{k=1}^K \widehat w_k \frac{1}{\sum_{i=1}^n \eta_i I(S_i = 0)} \sum_{i=1}^n \frac{(1- \widetilde p_k(X^{(k)}_i)) I(S_i \in \mathcal{S}^k, A_i= a)}{\widehat e_{a,k}(X_i^{(k)}) \widetilde p_k(X^{(k)}_i)} Y_i.
\end{equation}
Here, the estimator $\widetilde p_k(X^{(k)})$ for $\Pr_k[S \in \mathcal{S}^k|X^{(k)}]$ uses both data from the trials and target population and needs to account for the weighted sampling design. In our analysis we used a weighted logistic regression model with weights equal to one for NLST and PLCO participants and the survey sampling weights for the NHANES participants. The estimator $\widehat e_{a,k}(X^{(k)})$ is an estimator for $\Pr_k[A =a|X^{(k)}, S \in \mathcal{S}^k]$ and is only fit using trial data and does therefore not need to be modified to account for the sampling weights in the target population data.  

Last, we modify the doubly robust estimator as follows:
\begin{align*}
\widehat \psi^\eta_{\text{\tiny DR}}(a) = \sum_{k=1}^K \widehat w_k &\frac{1}{\sum_{i=1}^n \eta_iI(S_i = 0)} \\& \times \sum_{i=1}^n \Bigg( \frac{(1- \widetilde p_k(X^{(k)}_i)) I(S_i \in \mathcal{S}^k, A_i= a)}{\widehat e_{a,k}(X_i^{(k)}) \widetilde p_k(X^{(k)}_i)} (Y_i - \widehat g_{a,k}(X^{(k)}_i)) + \eta_i I(S_i = 0) \widehat g_{a,k}(X^{(k)}_i) \Bigg).
\end{align*}
We used the bootstrap for variance estimation and confidence interval (CI) construction. We accounted for the NHANES complex sampling design by using a stratified bootstrap \cite{shao2003impact,rao1988resampling} where the resampling is done so it is consistent with the NHANES sampling design (i.e.,~the resampling is done at the primary sampling unit and strata level \cite{curtin2012national}). For the naive pooled estimator, we used the non-parametric bootstrap stratified by trial. 

\subsection{Results}
We estimated the six year risk of being diagnosed with lung cancer to be $0.073$ (95\% bootstrap interval $[0.065, 0.082]$) using the g-formula estimator; $0.072$ (95\% bootstrap interval $[0.065, 0.079]$) using the weighting estimator; and $0.072$ (95\% bootstrap interval $[0.064, 0.080]$) using the doubly robust estimator. For comparison, the naive pooled estimator using only the trial data produced an estimated risk of $0.050$ (95\% bootstrap interval $[0.048, 0.053]$). Thus, all three estimators for the potential outcome mean in the target population that we proposed produced very similar point estimates that  were substantially larger compared with the naive pooled estimator. In fact, the bootstrap interval for the naive pooled estimator did not overlap with the bootstrap interval for the other three estimators. This may be due to differences in the distribution of education between the trials and the target population (see Table \ref{tab:DA-app} in Supplementary Web Appendix \ref{app:da} for the distribution of education levels) and higher education levels being associated with lower risk of lung cancer diagnosis (see Table \ref{tab-da1-app} in Supplementary Web Appendix \ref{app:da} for prevalence rates by education level).  


\section{Discussion}
\label{sec:Dis}

We provided identifiability results and proposed three estimators for the potential outcome mean in the target population for causally interpretable meta-analysis with systematically missing data. We studied the large-sample properties of the estimators and illustrated that the estimators have good finite-sample performance in simulation studies. Last, we estimated the risk of lung cancer diagnosis when chest radiography screening is applied using data from two large randomized controlled trials and using target population data from the NHANES study.

Our methods rely on the untestable missing at random assumption (Assumption A6), it would be of interest to develop sensitivity analysis methods to evaluate how violations of that assumption may affect the results \cite{dahabreh2019sensitivity}. We only considered the setting where the only source of missing data was systematic missing data and combining the methods we develop here with methods for handling within-trial or within-target missing data is of interest. Future work could also consider extensions of the methods to address failure time outcomes and covariate measurement error.

Throughout, we focused on meta-analyses that aim to estimate causal estimands such as potential outcome means and average treatment effects in the target population. In related work, Kundu et al.\cite{kundu2019generalized} developed a generalized method of moments approach for combining information from multiple parametric regression models to estimate a regression coefficient in the presence of systematically missing data. Their approach rests on the assumption that the joint distribution of the outcome and covariates is the same across all the studies being combined, and requires data from a ``reference sample'' that can be used to estimate the joint distribution of the covariates. It would be interesting to explore whether the approach we used for causal estimands can be extended to meta-analyses of regression models. 

The estimators we proposed combine data from all trials within a given missingness pattern to estimate the potential outcome mean in the target population and then form a convex combination of the pattern-specific estimates to construct a summary estimate. An alternative approach is to ``transport'' each trial separately and then use our methods to calculate the summary estimate combining the trial-specific estimators \cite{vo2019rethinking}. If that approach is taken, the methods we propose here can use robust and efficient trial-specific estimators \cite{dahabreh2019efficient}, combining their estimates using optimal weights, and accounting for the correlation induced by standardization to the same target population. In other words, transporting each trial separately can be viewed as a special case of our methods when each trial collects different covariates. For instance, this kind of analysis may be particularly attractive in multi-cohort observational analyses where the contributing cohorts have different data structures. In addition, working with each trial separately may be appealing even when some trials have the same missingness pattern when, for each trial, either the outcome, or both the trial participation and treatment models can be correctly specified, whereas correctly specifying at least one group of models for the pooled data from all trials with a given missingness pattern is more challenging. We note, however, that separately transporting each trial requires stronger positivity conditions because conditions A1 and A5 need to hold for each trial rather than the aggregate of all trials with the same missingness pattern.

\clearpage
\bibliographystyle{unsrt}
\bibliography{References}

\newpage 
\appendix 

\renewcommand{\theequation}{A.\arabic{equation}}
\setcounter{equation}{0}

\section{Proofs}
\subsection{Proof of identifiability results}
\label{Proof-ID}
\begin{proof}
Assumptions A1 through A6 imply that for each $k \in \{1, \ldots, K\}$ we can write the potential outcome mean in the target population $\E[Y^a|S=0]$ as
\begin{align*}
\E[Y^a|S=0] &= \E[\E[Y^a|X, S=0]|S=0] \\
&= \E_k[\E_k[Y^a|X, S=0]|S=0] \\
&= \E_k[\E_k[Y^a|X, S \in \mathcal{S}^k]|S=0] \\
&= \E_k[\E_k[Y^a|X, S \in \mathcal{S}^k, A=a]|S=0] \\
&= \E_k[\E_k[Y|X, S \in \mathcal{S}^k, A=a]|S=0] \\
&= \E_k[\E_k[Y|X^{(k)}, S \in \mathcal{S}^k, A=a]|S=0]\\
&=\psi_k(a),
\end{align*}
where all expectations above are well defined under positivity conditions A3 and A5. 

From the above result, and the constraint $\sum_{k=1}^K w_k^* =1$, it follows that
\begin{align*}
 \psi(a)  = \sum_{k=1}^K w_k^* \psi_k(a) = \sum_{k=1}^K w_k^* \E_k[\E_k[Y|X^{(k)}, A=a, S \in \mathcal{S}^k]|S=0] = \E[Y^a|S=0].
\end{align*}

\clearpage
Turning our attention to the weighting re-expression of the identifiability result, for each $k \in \{ 1, \ldots, K \}$, we have
\begin{align*}
\psi_k(a)  &= \E_k[\E_k[Y|X^{(k)}, A=a, S \in \mathcal{S}^k]|S=0] \\ 
\quad & = \E_k\left[\E_k\left[\frac{Y I(A=a, S \in \mathcal{S}^k)}{\Pr_k[A=a|X^{(k)}, S \in \mathcal{S}^k] \Pr_k[S \in \mathcal{S}^k|X^{(k)}]}\Bigg|X^{(k)}\right]\Bigg|S=0\right] \\
\quad & = \E_k\left[\frac{I(S=0)}{\Pr_k[S=0]}\E_k\left[\frac{YI(A=a, S \in \mathcal{S}^k)}{\Pr_k[A=a|X^{(k)}, S \in \mathcal{S}^k] \Pr_k[S \in \mathcal{S}^k|X^{(k)}]}\Bigg|X^{(k)}\right]\right] \\
\quad & = \frac{1}{\Pr_k[S=0]}\E_k\left[\E_k\left[\frac{Y I(A=a, S \in \mathcal{S}^k) (1 - \Pr_k[S \in \mathcal{S}^k|X^{(k)}])}{\Pr_k[A=a|X^{(k)}, S \in \mathcal{S}^k] \Pr_k[S \in \mathcal{S}^k|X^{(k)}]}\Bigg|X^{(k)}\right]\right] \\
\quad & = \frac{1}{\Pr_k[S=0]}\E_k\left[\frac{Y I(A=a, S \in \mathcal{S}^k) \Pr_k[S = 0|X^{(k)}]}{\Pr_k[A=a|X^{(k)}, S \in \mathcal{S}^k] \Pr_k[S \in \mathcal{S}^k|X^{(k)}]}\right].
\end{align*}
\end{proof}

Thus, it follows that $$ \psi(a) = \sum_{k=1}^K w_k^*\frac{1}{\Pr_k[S=0]}\E_k\left[\frac{I(A=a, S \in \mathcal{S}^k) \Pr_k[S = 0|X^{(k)}] Y}{\Pr_k[A=a|X^{(k)}, S \in \mathcal{S}^k] \Pr_k[S \in \mathcal{S}^k|X^{(k)}]}\right].$$

\subsection{First Order Influence Function}
\label{Inf-Func}
We will use pathwise derivatives to calculate the first order influence function of $\psi(a)$ using the identifiability result in expression \eqref{Out-Id} (see e.g.,~\cite{van2000asymptotic} for more details on influcence function calculations). Let $p_t$ denote a one dimensional submodel with $t \in [0,1[$ and $t=0$ corresponds to the true data law. Using that differentiation is a linear operator
\begin{align*}
\frac{\partial \psi_{p_t}(a)}{\partial t} \bigg|_{t=0} = \sum_{k=1}^K w_k \frac{\partial }{\partial t}\E_{k, p_t}[\E_{k, p_t}[Y|X^{(k)}, A=a, I(S \in \mathcal{S}^k)=1]|S=0]\Bigg|_{t=0}.
\end{align*}
Result in Appendix D in \cite{dahabreh2019efficient} show that 
\begin{align*}
&\frac{\partial }{\partial t}\E_{k,p_t}[\E_{k, p_t}[Y|X^{(k)}, A=a, I(S \in \mathcal{S}^k)=1]|S=0]\Bigg|_{t=0} \\ &\quad = \E_{k, p_0} \Bigg[ \dfrac{1}{\Pr_{k, p_0}[S = 0]}  \Bigg\{ I(S=0) \Big\{\E_{k, p_0}[Y | X^{(k)}, I(S \in \mathcal{S}^k) =1, A = a] - \psi_{p_0}(a)\Big\} \\
      &\quad\quad\quad\quad+ \dfrac{I(S \in \mathcal{S}^k, A = a) \Pr_{k, p_0}[S = 0 | X^{(k)}]}{\Pr_{k, p_0}[S \in \mathcal{S}^k|X^{(k)}] \Pr_{k, p_0}[A = a| X^{k}, I(S \in \mathcal{S}^k) = 1] } \\
      & \quad \quad \quad \quad\quad\quad\quad\times \Big\{ Y - \E_{k, p_0}[Y | X^{(k)}, I(S \in \mathcal{S}^k) = 1, A = a]\Big\} \Bigg\} u(O) \Bigg],
 \end{align*}
 where $u(O)$ is the score of the observable data. It follows that for any set of weights $\{w_1, \ldots, w_K\}$ the influence function of $\psi(a)$ under the non-parametric model is
 \begin{align}
IF = &\sum_{k=1}^K w_k \dfrac{1}{\Pr_{k, p_0}[S = 0]}  \Bigg\{ I(S=0) \Big\{ \E_{k, p_0}[Y | X^{(k)}, I(S \in \mathcal{S}^k) =1, A = a] - \psi_{p_0}(a)\Big\} \nonumber \\
      &\quad+ \dfrac{I(S \in \mathcal{S}^k, A = a) \Pr_{k, p_0}[S = 0 | X^{(k)}]}{\Pr_{k, p_0}[S \in \mathcal{S}^k|X^{(k)}] \Pr_{k, p_0}[A = a| X^{k}, I(S \in \mathcal{S}^k) = 1] } \nonumber \\
      & \quad \quad \quad \times \Big\{ Y - \E_{k, p_0}[Y | X^{(k)}, I(S \in \mathcal{S}^k) = 1, A = a]\Big\} \Bigg\}. \label{Inf-Func-Form}
 \end{align}


\subsection{Proof of Theorem \ref{DR-thm}}

As $n \rightarrow \infty$ we have 
\begin{align*} 
\widehat \psi_{\text{\tiny DR}}(a) \overset{P}{\longrightarrow} &\sum_{k=1}^K w^*_k \Bigg( \frac{1}{\Pr_k[S=0]} \E_k\Bigg[ \frac{(1- p^*_k(X^{(k)})) I(S \in \mathcal{S}^k, A= a)}{e_{a,k}^*(X^{(k)}) p^*_k(X^{(k)})} \big(Y - g_{a,k}^*(X^{(k)})\big) \\ 
&\quad \quad \quad \quad \quad + I(S = 0) g_{a,k}^*(X^{(k)}) \Bigg]\Bigg).
\end{align*}
We will now show that the right hand side of the above equation is equal to $\psi(a)$ for the two cases listed in Assumption B1.
\\
\\
Case 1: Let $k \in \{1, \ldots, K\}$ be given and assume $\widehat g_{a,k}(X^{(k)}) \overset{P}{\longrightarrow} \E_k[Y|X^{(k)}, S \in \mathcal{S}^k, A=a]$. We don't assume that either of the estimators $\widehat p_k(X^{(k)})$ or $\widehat e_{a,k}(X^{(k)})$ are consistent. Under this assumptions we have
\begin{align*}
\frac{1}{\Pr_k[S=0]}\E_k\left[I(S=0) g^*_a(X^{(k)})\right] &= \E_k[\E_k[Y|X^{(k)}, S \in \mathcal{S}^k, A=a]|S=0] \\
&= \E[Y^a|S=0],
\end{align*}
and
\begin{align*}
&\E_k\Bigg[\frac{(1- p^*_k(X^{(k)})) I(S \in \mathcal{S}^k, A= a)}{e_{a,k}^*(X^{(k)}) p^*_k(X^{(k)})} \big(Y - g_{a,k}^*(X^{(k)})\big) \Bigg] \\ &= \E_k\Bigg[\E_k\Bigg[\frac{(1- p^*_k(X^{(k)})) I(S \in \mathcal{S}^k, A= a)}{e_{a,k}^*(X^{(k)}) p^*_k(X^{(k)})} \big(Y - \E_k[Y|X^{(k)}, S \in \mathcal{S}^k, A=a]\big)\Bigg| X^{(k)}\Bigg] \Bigg] \\
&=  \E_k\Bigg[\frac{(1- p^*_k(X^{(k)})) \Pr_k[S \in \mathcal{S}^k, A= a|X^{(k)}]}{e_{a,k}^*(X^{(k)}) p^*_k(X^{(k)})}  \big(Y - \E_k[Y|X^{(k)}, S \in \mathcal{S}^k, A=a]\big)\Bigg| X^{(k)}, S \in \mathcal{S}^k, A=a\Bigg] \\
&=0
\end{align*}
\\
\\
Case 2: Let $k \in \{1, \ldots, K\}$ be given, and assume $\widehat p_k(X^{(k)}) \overset{P}{\longrightarrow} \Pr_k[S \in \mathcal{S}^k|X^{(k)}]$ and $\widehat e_{a,k}(X^{(k)}) \overset{P}{\longrightarrow} \Pr_k[A =a|X^{(k)}, S \in \mathcal{S}^k]$. We don't assume that $\widehat g_{a,k}(X^{(k)})$ is a consistent estimator. Under this assumption
\[
\frac{1}{\Pr_k[S=0]}\E_k\left[I(S=0) g^*_a(X^{(k)})\right] = \frac{1}{\Pr_k[S=0]}\E_k\left[{\Pr}_k[S=0|X^{(k)}] g^*_a(X^{(k)})\right]
\]
and 
\begin{align*}
\frac{1}{\Pr_k[S=0]} \E_k\Bigg[ &\frac{(1- p^*_k(X^{(k)})) I(S \in \mathcal{S}^k, A= a)}{e_{a,k}^*(X^{(k)}) p^*_k(X^{(k)})} g_{a,k}^*(X^{(k)}) \Bigg]  \\
&= \frac{1}{\Pr_k[S=0]} \E_k\Bigg[ \frac{(1- p^*_k(X^{(k)}))g_{a,k}^*(X^{(k)})}{e_{a,k}^*(X^{(k)}) p^*_k(X^{(k)})} E_k[I(S \in \mathcal{S}^k, A= a)|X^{(k)}] \Bigg] \\
&=\frac{1}{\Pr_k[S=0]}\E_k\left[{\Pr}_k[S=0|X^{(k)}] g^*_a(X^{(k)})\right].
\end{align*}
We have
\begin{align*}
\frac{1}{\Pr_k[S=0]} \E_k\Bigg[ &\frac{(1- p^*_k(X^{(k)})) I(S \in \mathcal{S}^k, A= a)}{e_{a,k}^*(X^{(k)}) p^*_k(X^{(k)})} Y \Bigg] \\
&= \frac{1}{\Pr_k[S=0]} \E_k\big[\Pr_{k}[S=0|X^{(k)}] \E_k[Y|X^{(k)}, A=a, S \in \mathcal{S}^k] \big] \\
&= \frac{1}{\Pr_k[S=0]} \E_k\big[I(S=0) \E_k[Y|X^{(k)}, A=a, S \in \mathcal{S}^k] \big] \\
&= \E_k[\E_k[Y|X^{(k)}, A=a, S \in \mathcal{S}^k] |S=0] \\
&= \E[Y^a|S=0]\\
&= \psi(a).
\end{align*}
Combing the above, we have shown that for both cases and each $k \in \{1, \ldots, K\}$, 
\[
\frac{1}{\Pr_k[S=0]} \E_k\Bigg[ \frac{(1- p^*_k(X^{(k)})) I(S \in \mathcal{S}^k, A= a)}{e_{a,k}^*(X^{(k)}) p^*_k(X^{(k)})} \big(Y - g_{a,k}^*(X^{(k)})\big) + I(S = 0) g_{a,k}^*(X^{(k)}) \Bigg] = \psi(a).
\]
As the weights sum to one we get
\[
\widehat \psi_{\text{\tiny DR}}(a) \overset{P}{\longrightarrow} \sum_{k=1}^K w^*_k \psi(a) = \psi(a),
\]
completing the consistency proof.

Now consider the asymptotic representation given by equation \eqref{As-Rep}. Rewrite 
\begin{align*}
\sqrt{n}\big(\widehat \psi_{\text{\tiny DR}}(a) - \psi(a)\big) 
&= \sum_{k=1}^K w_k^* \mathbb{G}_{k}\bigg(L(\widehat g_{a,k}(X^{(k)}), \widehat e_{a,k}(X^{(k)}),\widehat \gamma_{k}, \widehat p_k(X^{(k)}))\bigg)
\\ &- \sum_{k=1}^K w_k^* \mathbb{G}_{k}\bigg(L(g_{a,k}^*(X^{(k)}), e_{a,k}^*(X^{(k)}),\gamma_{k}, p^*_k(X^{(k)}))\bigg) \\
&+ \sum_{k=1}^K w_k^* \mathbb{G}_{k}\bigg(L(g_{a,k}^*(X^{(k)}), e_{a,k}^*(X^{(k)}),\gamma_{k}, p^*_k(X^{(k)}))\bigg) \\
&+ \sqrt{n} \left( \sum_{k=1}^K w_k^* \E\bigg[L(\widehat g_{a,k}(X^{(k)}), \widehat e_{a,k}(X^{(k)}),\widehat \gamma_{k}, \widehat p_k(X^{(k)}))\bigg] - \sum_{k=1}^K w_k^* \psi(a) \right) \\
&+ \sqrt{n} \sum_{k=1}^K (\widehat w_k - w^*_k) \mathbb{P}_{k}\big(L(\widehat g_{a,k}(X^{(k)}), \widehat e_{a,k}(X^{(k)}),\widehat \gamma_k, \widehat p_k(X^{(k)}))\big)
\end{align*}
By the Donsker condition 
\begin{align*}
\mathbb{G}_{k}\bigg(&L(\widehat g_{a,k}(X^{(k)}), \widehat e_{a,k}(X^{(k)}),\widehat\gamma_k, \widehat p_k(X^{(k)}))\bigg) \\ &- \mathbb{G}_{k}\bigg(L(g_{a,k}^*(X^{(k)}), e_{a,k}^*(X^{(k)}),\gamma_{k}, p^*_k(X^{(k)}))\bigg) = o_p(1).
\end{align*}
Similar calculations to Appendix E in \cite{dahabreh2019efficient} show that for each $k \in \{1, \ldots, K\}$,
\begin{align}
E\big[w^*_k &L(\widehat g_{a,k}(X^{(k)}), \widehat e_{a,k}(X^{(k)}),\widehat \gamma_{k}, \widehat p_k(X^{(k)}))\big] -  w_k^* \psi(a) \nonumber \\
&\leq O_P\Bigg(\Big(\Big|\Big|\big(\widehat p_k(X^{(k)}) - {\Pr}_k[S \in \mathcal{S}^k|X^{(k)}] \Big|\Big|_2 + \Big|\Big|\widehat e_{a,k}(X^{(k)}) - {\Pr}_k[A =a|X^{(k)}, S \in \mathcal{S}^k]\Big|\Big|_2 \Big)\nonumber \\
&\times \Big|\Big| \widehat g_{a,k}(X^{(k)}) - \E_k[Y|X^{(k)}, S \in \mathcal{S}^k, A=a]\Big|\Big|_2 \Bigg), \label{as-rep-mean}
\end{align}
which completes the proof of the asymptotic representation. 

\subsection{Asymptotic representation of sample split doubly robust estimator}
\label{sec:Don}

As the consistency proof of Theorem \ref{DR-thm} does not require the Donsker condition, we only need to show that the asymptotic representation provided in Theorem \ref{DR-thm} holds for the sample split doubly robust estimator without requiring the Donsker assumption.

Recall that sample split doubly robust estimator is given by 
\[
\widehat \psi_{\text{\tiny DR}}^{(SS)}(a) = \frac{1}{2} \widehat \psi_{\text{\tiny DR}}^{(1)}(a) + \frac{1}{2}\widehat \psi_{\text{\tiny DR}}^{(2)}(a),
\]
where
\begin{equation*}
\widehat \psi_{\text{\tiny DR}}^{(1)}(a) = \sum_{k=1}^K \widehat w_k \frac{1}{\sum_{i \in \mathcal{B}_1} I(S_i = 0)} \sum_{i \in \mathcal{B}_1} \left( \widehat o_a^{(2)}(X^{(k)}_i, A_i,S_i) (Y_i - \widehat g_{a,k}^{(2)}(X^{(k)}_i)) + I(S_i = 0) \widehat g_{a,k}^{(2)}(X^{(k)}_i) \right)
\end{equation*}
and
\begin{equation*}
\widehat \psi_{\text{\tiny DR}}^{(2)}(a) = \sum_{k=1}^K \widehat w_k \frac{1}{\sum_{i \in \mathcal{B}_2} I(S_i = 0)} \sum_{i \in \mathcal{B}_2} \left( \widehat o_a^{(1)}(X^{(k)}_i, A_i,S_i) (Y_i - \widehat g_{a,k}^{(1)}(X^{(k)}_i)) + I(S_i = 0) \widehat g_{a,k}^{(1)}(X^{(k)}_i) \right).
\end{equation*}
For simplicity assume both set of indexes $\mathcal{B}_1$ and $\mathcal{B}_2$ are of equal size $m = \frac{n}{2}$. Define $\mathbb{P}_{k}^{(1)}(W)$ and $\mathbb{P}_{k}^{(2)}(W)$ as the analogs of $\mathbb{P}_{k}(W)$ where the average is taken over observations in $\mathcal{B}_1$ and $\mathcal{B}_2$, respectively. Define $\mathbb{G}_{k}^{(1)}(W) = \sqrt{m} (\mathbb{P}_{k}^{(1)}(W) - E[W])$ and $\mathbb{G}_{k}^{(2)}(W) = \sqrt{m} (\mathbb{P}_{k}^{(2)}(W) - E[W])$. For $j=1,2$ and $k \in \{1, \ldots, K\}$, let $\gamma_k^{(j)} = \Pr_k[S=0|\mathcal{B}_j]^{-1}$, $\widehat \gamma_k^{(1)} = \frac{\sum_{i \in \mathcal{B}_1} (I(S_i = 0) + I(S_i \in \mathcal{S}^{k}))}{\sum_{i \in \mathcal{B}_1} I(S_i = 0)}$ and $\widehat \gamma_k^{(2)} = \frac{\sum_{i \in \mathcal{B}_2} (I(S_i = 0) + I(S_i \in \mathcal{S}^{k}))}{\sum_{i \in \mathcal{B}_2} I(S_i = 0)}$. Rewrite 
\begin{align*}
\sqrt{n}\left(\widehat \psi_{\text{\tiny DR}}^{(SS)}(a) - \psi(a) \right) &= 
\frac{\sqrt{n}}{2} \left(\widehat \psi_{\text{\tiny DR}}^{(1)}(a) - \psi(a) \right) + \frac{\sqrt{n}}{2} \left(\widehat \psi_{\text{\tiny DR}}^{(2)}(a) - \psi(a) \right) 
\\ &= \frac{\sqrt{n}}{2} \sum_{k=1}^K (\widehat w_k - w^*_k) \mathbb{P}^{(1)}_{n,k}\big(L(\widehat g_{a,k}^{(2)}(X^{(k)}), \widehat e_{a,k}^{(2)}(X^{(k)}),\widehat \gamma_k^{(2)}, \widehat p^{(2)}_k(X^{(k)}))\big) 
\\ &+ \frac{\sqrt{n}}{2} \sum_{k=1}^K (\widehat w_k - w^*_k) \mathbb{P}^{(2)}_{n,k}\big(L(\widehat g_{a,k}^{(1)}(X^{(k)}), \widehat e_{a,k}^{(1)}(X^{(k)}),\widehat \gamma_k^{(1)}, \widehat p^{(1)}_k(X^{(k)}))\big) \\
&+ \frac{\sqrt{n}}{2}\left(\sum_{k=1}^K w_k^* \mathbb{P}^{(1)}_{n,k}\big(L(\widehat g_{a,k}^{(2)}(X^{(k)}), \widehat e_{a,k}^{(2)}
(X^{(k)}),\widehat \gamma_k^{(2)}
, \widehat p^{(2)}_k(X^{(k)}))\big) - \psi(a) \right) \\
&+ \frac{\sqrt{n}}{2}  \left( \sum_{k=1}^K w_k^* \mathbb{P}^{(2)}_{n,k}\big(L(\widehat g_{a,k}^{(1)}(X^{(k)}), \widehat e_{a,k}^{(1)}
(X^{(k)}),\widehat \gamma_k^{(1)}
, \widehat p^{(1)}_k(X^{(k)}))\big) - \psi(a) \right) \\
&= \frac{1}{\sqrt{2}} \Bigg( \sum_{k=1}^K w_k^* \bigg(\mathbb{G}^{(1)}_{n,k}\big(L(\widehat g_{a,k}^{(2)}(X^{(k)}), \widehat e_{a,k}^{(2)}
(X^{(k)}),\widehat \gamma_k^{(2)}
, \widehat p^{(2)}_k(X^{(k)}))\big) \\
& - \mathbb{G}^{(1)}_{n,k}\big(L(g_{a,k}^{*}(X^{(k)}), e_{a,k}^{*}
(X^{(k)}),\gamma_k
,  p^{*}(X^{(k)}))\big)\bigg) \\
&   + \sum_{k=1}^K w_k^* \mathbb{G}^{(1)}_{n,k}\big(L( g_{a,k}^{*}(X^{(k)}), e_{a,k}^{*}
(X^{(k)}),\gamma_k
,  p^{*}(X^{(k)}))\big) \\
&+ \sqrt{m} \left(\E\left[ \sum_{k=1}^K w_k^* L(\widehat g_{a,k}^{(2)}(X^{(k)}), \widehat e_{a,k}^{(2)}
(X^{(k)}),\widehat \gamma_k^{(2)}
, \widehat p^{(2)}_k(X^{(k)}))\right] - \psi(a)\right)\Bigg)  \\
&+  \frac{1}{\sqrt{2}} \Bigg( \sum_{k=1}^K w_k^* \bigg(\mathbb{G}^{(2)}_{n,k}\big(L(\widehat g_{a,k}^{(1)}(X^{(k)}), \widehat e_{a,k}^{(1)}
(X^{(k)}),\widehat \gamma_k^{(1)}
, \widehat p^{(1)}_k(X^{(k)}))\big) \\
& - \mathbb{G}^{(2)}_{n,k}\big(L(g_{a,k}^{*}(X^{(k)}), e_{a,k}^{*}
(X^{(k)}),\gamma_k
, p^{*}(X^{(k)}))\big)\bigg) \\
&   + \sum_{k=1}^K w_k^* \mathbb{G}^{(2)}_{n,k}\big(L( g_{a,k}^{*}(X^{(k)}), e_{a,k}^{*}
(X^{(k)}), \gamma_k
,  p^{*}(X^{(k)}))\big) \\
&+ \sqrt{m} \left(\E\left[\sum_{k=1}^K w_k^*  L(\widehat g_{a,k}^{(1)}(X^{(k)}), \widehat e_{a,k}^{(1)}
(X^{(k)}),\widehat \gamma_k^{(1)}
, \widehat p^{(1)}_k(X^{(k)}))\right] - \psi(a)\right) \Bigg) \\ &+ O_P\left(\sqrt{n} \sum_{k=1}^K (\widehat w_k - w_k^*)\right).
\end{align*}
By rearranging terms we get
\begin{align*}
&\sqrt{n}\left(\widehat \psi_{\text{\tiny DR}}^{(SS)}(a) - \psi(a) \right) \\
&= \frac{1}{\sqrt{2}} \Big( \sum_{k=1}^K w_k^* \mathbb{G}^{(1)}_{n,k}\big(L( g_{a,k}^{*}(X^{(k)}), e_{a,k}^{*}
(X^{(k)}),\gamma_k
,  p^{*}(X^{(k)}))\big) \\ 
& \quad \quad+  \sum_{k=1}^K w_k^* \mathbb{G}^{(2)}_{n,k}\big(L( g_{a,k}^{*}(X^{(k)}), e_{a,k}^{*}
(X^{(k)}),\gamma_k
,  p^{*}(X^{(k)}))\big) \Big) \\
&+ \frac{1}{\sqrt{2}} \Bigg( \sum_{k=1}^K w_k^* \bigg(\mathbb{G}^{(1)}_{n,k}\big(L(\widehat g_{a,k}^{(2)}(X^{(k)}), \widehat e_{a,k}^{(2)}
(X^{(k)}),\widehat \gamma_k^{(2)}
, \widehat p^{(2)}_k(X^{(k)}))\big) \\
& - \mathbb{G}^{(1)}_{n,k}\big(L(g_{a,k}^{*}(X^{(k)}), e_{a,k}^{*}
(X^{(k)}),\gamma_k
,  p^{*}(X^{(k)}))\big) \\
&+ \sum_{k=1}^K w_k^* \bigg(\mathbb{G}^{(2)}_{n,k}\big(L(\widehat g_{a,k}^{(1)}(X^{(k)}), \widehat e_{a,k}^{(1)}
(X^{(k)}),\widehat \gamma_k^{(1)}
, \widehat p^{(1)}_k(X^{(k)}))\big) \\
& - \mathbb{G}^{(2)}_{n,k}\big(L(g_{a,k}^{*}(X^{(k)}), e_{a,k}^{*}
(X^{(k)}),\gamma_k
,  p^{*}(X^{(k)}))\big) \bigg) \Bigg) \\
&+ \frac{1}{\sqrt{2}} \sqrt{m} \left(\E\left[ \sum_{k=1}^K w_k^*L(\widehat g_{a,k}^{(2)}(X^{(k)}), \widehat e_{a,k}^{(2)}
(X^{(k)}),\widehat \gamma_k^{(1)}
, \widehat p^{(2)}_k(X^{(k)}))\right] - \psi(a)\right) \\\
&+  \frac{1}{\sqrt{2}} \sqrt{m} \left(\E\left[ \sum_{k=1}^K w_k^*L(\widehat g_{a,k}^{(1)}(X^{(k)}), \widehat e_{a,k}^{(1)}
(X^{(k)}),\widehat \gamma_k^{(1)}
, \widehat p^{(1)}_k(X^{(k)}))\right] - \psi(a)\right) 
\\ &+ O_P\left(\sqrt{n} \sum_{k=1}^K (\widehat w_k - w_k^*)\right).
\end{align*}
We have
\begin{align*}
&\mathbb{P}^{(1)}_{n,k}\big(L( g_{a,k}^{*}(X^{(k)}), e_{a,k}^{*}
(X^{(k)}),\gamma_k
,  p^{*}(X^{(k)}))\big) + \mathbb{P}^{(2)}_{n,k}\big(L( g_{a,k}^{*}(X^{(k)}), e_{a,k}^{*}
(X^{(k)}),\gamma_k
,  p^{*}(X^{(k)}))\big)  \\
&= \frac{2}{n} \sum_{i \in \mathcal{B}_1} L( g_{a,k}^{*}(X_i^{(k)}), e_{a,k}^{*}
(X_i^{(k)}),\gamma_k
,  p^{*}(X_i^{(k)})) + \frac{2}{n} \sum_{i \in \mathcal{B}_2} L( g_{a,k}^{*}(X_i^{(k)}), e_{a,k}^{*}
(X_i^{(k)}),\gamma_k
,  p^{*}(X_i^{(k)})) \\
&= 2 \mathbb{P}_{k} \big(L( g_{a,k}^{*}(X^{(k)}), e_{a,k}^{*}
(X^{(k)}),\gamma_k
,  p^{*}(X^{(k)}))\big). 
\end{align*}
Using the result above
\begin{align*}
&\frac{1}{\sqrt{2}} \Big(\mathbb{P}^{(1)}_{n,k}\big(L( g_{a,k}^{*}(X^{(k)}), e_{a,k}^{*}
(X^{(k)}),\gamma_k
,  p^{*}(X^{(k)}))\big) - \E\left[L( g_{a,k}^{*}(X^{(k)}), e_{a,k}^{*}
(X^{(k)}),\gamma_k
,  p^{*}(X^{(k)}))\right] \\
&+ \mathbb{P}^{(2)}_{n,k}\big(L( g_{a,k}^{*}(X^{(k)}), e_{a,k}^{*}
(X^{(k)}),\gamma_k
,  p^{*}(X^{(k)}))\big) - \E\left[L( g_{a,k}^{*}(X^{(k)}), e_{a,k}^{*}
(X^{(k)}),\gamma_k
,  p^{*}(X^{(k)}))\right]\Big) \\
&= \sqrt{2} \mathbb{P}_{k}\big( L( g_{a,k}^{*}(X^{(k)}), e_{a,k}^{*}
(X^{(k)}),\gamma_k
,  p^{*}(X^{(k)})) \big)- \sqrt{2} \E\left[L( g_{a,k}^{*}(X^{(k)}), e_{a,k}^{*}
(X^{(k)}),\gamma_k
,  p^{*}(X^{(k)}))\right].
\end{align*}
It follows that
\begin{align*}
&\frac{1}{\sqrt{2}} \left( \sum_{k=1}^K w_k^* \mathbb{G}^{(1)}_{n,k}\big(L( g_{a,k}^{*}(X^{(k)}), e_{a,k}^{*}
(X^{(k)}),\gamma_k
,  p^{*}(X^{(k)}))\big)+  \sum_{k=1}^K w_k^* \mathbb{G}^{(2)}_{n,k}\big(L( g_{a,k}^{*}(X^{(k)}), e_{a,k}^{*}
(X^{(k)}),\gamma_k
,  p^{*}(X^{(k)}))\big) \right)\\
&= \frac{\sqrt{m}}{\sqrt{2}} \Bigg( \sum_{k=1}^K w_k^* \mathbb{P}^{(1)}_{n,k}\big(L( g_{a,k}^{*}(X^{(k)}), e_{a,k}^{*}
(X^{(k)}),\gamma_k
,  p^{*}(X^{(k)}))\big) - \sum_{k=1}^K w_k^* \E\left[L( g_{a,k}^{*}(X^{(k)}), e_{a,k}^{*}
(X^{(k)}),\gamma_k
,  p^{*}(X^{(k)}))\right] \Bigg)\\
&+ \frac{\sqrt{m}}{\sqrt{2}} \Bigg( \sum_{k=1}^K w_k^* \mathbb{P}^{(2)}_{n,k}\big(L( g_{a,k}^{*}(X^{(k)}), e_{a,k}^{*}
(X^{(k)}),\gamma_k
,  p^{*}(X^{(k)}))\big) - \sum_{k=1}^K w_k^* \E\left[L( g_{a,k}^{*}(X^{(k)}), e_{a,k}^{*}
(X^{(k)}),\gamma_k
,  p^{*}(X^{(k)}))\right] \Bigg) \\
&= \sqrt{n}\left( \sum_{k=1}^K w_k^* \mathbb{P}_{k}\big( L( g_{a,k}^{*}(X^{(k)}), e_{a,k}^{*}
(X^{(k)}),\gamma_k
,  p^{*}(X^{(k)}))\big) -\sum_{k=1}^K w_k^* \E\left[L( g_{a,k}^{*}(X^{(k)}), e_{a,k}^{*}
(X^{(k)}),\gamma_k
,  p^{*}(X^{(k)}))\right]\right) \\
&=  \sum_{k=1}^K w_k^* \mathbb{G}_{k}\big( L( g_{a,k}^{*}(X^{(k)}), e_{a,k}^{*}
(X^{(k)}),\gamma_k
,  p^{*}(X^{(k)}))\big).
\end{align*}
For each $k \in \{1, \ldots, K\}$ the inequality in \eqref{as-rep-mean} and that $n = 2m$ gives
\begin{align*}
&\frac{1}{\sqrt{2}} \sqrt{m} \left(\E\left[ \mathbb{P}^{(1)}_{n,k}\big(L(\widehat g_{a,k}^{(2)}(X^{(k)}), \widehat e_{a,k}^{(2)}
(X^{(k)}),\widehat \gamma_k^{(1)}
, \widehat p^{(2)}_k(X^{(k)}))\big)\right] - \psi(a)\right) \\\
&+  \frac{1}{\sqrt{2}} \sqrt{m} \left(\E\left[ \mathbb{P}^{(2)}_{n,k}\big(L(\widehat g_{a,k}^{(1)}(X^{(k)}), \widehat e_{a,k}^{(1)}
(X^{(k)}),\widehat \gamma_k^{(1)}
, \widehat p^{(1)}_k(X^{(k)}))\big)\right] - \psi(a)\right) \\
&\leq \sqrt{n} O_P\Bigg(\Big(\Big|\Big|\big(\widehat p_k(X^{(k)}) - \Pr_k[S \in \mathcal{S}^k|X^{(k)}] \Big|\Big|_2 + \Big|\Big|\widehat e_{a,k}(X^{(k)}) - \Pr_k[A =a|X^{(k)}, S \in \mathcal{S}^k]\Big|\Big|_2 \Big)\\
&\times \Big|\Big| \widehat g_{a,k}(X^{(k)}) - \E_k[Y|X^{(k)}, S \in \mathcal{S}^k, A=a]\Big|\Big|_2 \Bigg).
\end{align*}
For each $k \in \{1, \ldots, K\}$ define
\begin{align*}
C_{1,k} = &\mathbb{G}^{(1)}_{n,k}\big(L(\widehat g_{a,k}^{(2)}(X^{(k)}), \widehat e_{a,k}^{(2)}
(X^{(k)}),\widehat \gamma_k^{(2)}
, \widehat p^{(2)}_k(X^{(k)}))\big) - \mathbb{G}^{(1)}_{n,k}\big(L(g_{a,k}^{*}(X^{(k)}), e_{a,k}^{*}
(X^{(k)}),\gamma_k
,  p^{*}(X^{(k)}))\big)    \\
&= \sqrt{m} \frac{1}{m} \sum_{i \in \mathcal{B}_1} \Bigg( L(\widehat g_{a,k}^{(2)}(X_i^{(k)}), \widehat e_{a,k}^{(2)} (X_i^{(k)}),\widehat \gamma_k^{(2)}
, \widehat p^{(2)}_k(X_i^{(k)}))- L(g_{a,k}^*(X_i^{(k)}),  e_{a,k}^*(X_i^{(k)}), \gamma_k,  p^*_k(X_i^{(k)})) \\
&- \left( \E \left[ L(\widehat g_{a,k}^{(2)}(X_i^{(k)}), \widehat e_{a,k}^{(2)}
(X_i^{(k)}),\widehat \gamma_k^{(2)}
, \widehat p^{(2)}_k(X_i^{(k)})) \right]  -\E\left[ L(g_{a,k}^*(X^{(k)}),  e_{a,k}^*
(X^{(k)}), \gamma_k
,  p^*_k(X^{(k)})) \right]\right) \Bigg)
\end{align*}
Define $\mathcal{O}_1 = \{O_i, i \in \mathcal{B}_1\}$ and $\mathcal{O}_2 = \{O_i, i \in \mathcal{B}_2\}$. As $\mathcal{O}_1$ and $\mathcal{O}_2$ are independent, for any $i \in \mathcal{B}_1$
\begin{align*}
 &\E[L(\widehat g_{a,k}^{(2)}(X_i^{(k)}), \widehat e_{a,k}^{(2)} (X_i^{(k)}),\widehat \gamma_k^{(2)}
, \widehat p^{(2)}_k(X_i^{(k)}))- L(g_{a,k}^*(X_i^{(k)}),  e_{a,k}^*(X_i^{(k)}), \gamma_k,  p^*_k(X_i^{(k)}))|\mathcal{O}_2] \\
&= \E[L(\widehat g_{a,k}^{(2)}(X_i^{(k)}), \widehat e_{a,k}^{(2)} (X_i^{(k)}),\widehat \gamma_k^{(2)}
, \widehat p^{(2)}_k(X_i^{(k)}))- L(g_{a,k}^*(X_i^{(k)}),  e_{a,k}^*(X_i^{(k)}), \gamma_k,  p^*_k(X_i^{(k)}))]
\end{align*}
Hence $\E[C_{1,k}|\mathcal{O}_2] = 0.$ Again, using that $\mathcal{O}_1$ and $\mathcal{O}_2$ are independent we have
\begin{align*}
\text{Var}[C_{1,k}|\mathcal{O}_2] &= \text{Var}\left[  L(\widehat g_{a,k}^{(2)}(X_i^{(k)}), \widehat e_{a,k}^{(2)} (X_i^{(k)}),\widehat \gamma_k^{(2)}
, \widehat p^{(2)}_k(X_i^{(k)}))- L(g_{a,k}^*(X_i^{(k)}),  e_{a,k}^*(X_i^{(k)}), \gamma_k,  p^*_k(X_i^{(k)})) \bigg|\mathcal{O}_2\right],
\end{align*}
where $i$ is used to index a random observation with index in $\mathcal{B}_1$. Using this 
\begin{align*}
\text{Var}[C_{1,k}|\mathcal{O}_2] &\leq \Pr\left[\left( L(\widehat g_{a,k}^{(2)}(X_i^{(k)}), \widehat e_{a,k}^{(2)} (X_i^{(k)}),\widehat \gamma_k^{(2)}
, \widehat p^{(2)}_k(X_i^{(k)}))- L(g_{a,k}^*(X_i^{(k)}),  e_{a,k}^*(X_i^{(k)}), \gamma_k,  p^*_k(X_i^{(k)}))\right)^2\Bigg|\mathcal{O}_2\right].
\end{align*}
By assumptions B3, $\text{Var}[C_{1,k}|\mathcal{O}_2] = \Pr[C_{1,k}^2|\mathcal{O}_2] \rightarrow 0$ and by Chebyshev's inequality for any $\delta > 0$
\begin{align*}
\Pr[|C_{1,k}|  > \delta |\mathcal{O}_2] \leq \frac{\Pr[C_{1,k}^2|\mathcal{O}_2]}{\delta^2} \rightarrow 0 
\end{align*}
when $m \rightarrow \infty$. It follows that $\sum_{k=1}^K w_k^* C_{1,k} = o_p(1)$ and the same arguments with the roles of $\mathcal{B}_1$ and $\mathcal{B}_2$ switched shows that $\sum_{k=1}^K w_k^* C_{2,k} = o_p(1)$. Combining all the above gives 
\begin{align*}
\sqrt{n}\left(\widehat \psi_{\text{\tiny DR}}^{(SS)}(a) - \psi(a) \right) = \sum_{k=1}^K w_k^* \mathbb{G}_{k}\big( L( g_{a,k}^{*}(X^{(k)}), e_{a,k}^{*}
(X^{(k)}),\gamma_k
,  p^{*}(X^{(k)}))\big) + REM + o_p(1),
\end{align*}
where 
\begin{align*}
REM \leq &\sqrt{n} O_P\Bigg(\sum_{k=1}^K \Big|\Big| w_k^* - w^*_k\Big|\Big|_2+ \sum_{k=1}^K \Big|\Big| \widehat g_{a,k}(X^{(k)}) - \E_k[Y|X^{(k)}, S \in \mathcal{S}^k, A=a]\Big|\Big|_2  \\
&\times \Big( \Big|\Big|\widehat p_k(X^{(k)}) - {\Pr}_k[S \in \mathcal{S}^k|X^{(k)}] \Big|\Big|_2 + \Big|\Big|\widehat e_{a,k}(X^{(k)}) - {\Pr}_k[A =a|X^{(k)}, S \in \mathcal{S}^k]\Big|\Big|_2 \Big)\Bigg) .
\end{align*}

\subsection{Derivation of optimal weights}
\label{sec:opt-weights}
All expectations in this section are w.r.t.~the biased sampling model described in Section \ref{sec:Id}. The expected square of the influence function in expression \eqref{Inf-Func-Form} can be written as
\begin{align}
E[IF^2;w_1, \ldots, w_k] &= \sum_{k=1}^K w_k^2 \dfrac{1}{\Pr_{k}[S = 0]^2} \E_k\Bigg[ \Bigg( I(S=0) \Big\{ \E_{k}[Y | X^{(k)}, I(S \in \mathcal{S}^k) =1, A = a] - \psi(a)\Big\} \nonumber \\
      &\quad+ \dfrac{I(S \in \mathcal{S}^k, A = a) \Pr_{k}[S = 0 | X^{(k)}]}{\Pr_{k}[S \in \mathcal{S}^k|X^{(k)}] \Pr_{k}[A = a| X^{k}, I(S \in \mathcal{S}^k) = 1] } \nonumber \\
      & \quad \quad \quad \times \Big\{ Y - \E_{k}[Y | X^{(k)}, I(S \in \mathcal{S}^k) = 1, A = a]\Big\} \Bigg)^2 \Bigg] \nonumber \\
      &+2 \sum_{k=1}^K \sum_{j=1}^{k-1} w_k w_j \dfrac{1}{\Pr_{k}[S = 0] \Pr_{j}[S = 0]} \E\big[ (\E[Y | X^{(k)}, I(S \in \mathcal{S}^k) =1, A = a] - \psi(a)) \nonumber \\
      &\quad \quad \quad \quad \quad \quad \quad \quad \quad  \quad \quad \quad  \times  I(S=0)(\E[Y | X^{(j)}, I(S \in \mathcal{S}^j) =1, A = a] - \psi(a)) \big] \label{Inf-a}.
\end{align}
We use the notation $\E[IF^2;w_1, \ldots, w_k]$ to emphasize the dependence of the influence function on the weights $w_1, \ldots, w_k$. Define
\begin{align*}
V_k := &\dfrac{1}{\Pr_{k}[S = 0]^2} \E_k\Bigg[ \Bigg( I(S=0) \Big\{ \E_{k}[Y | X^{(k)}, I(S \in \mathcal{S}^k) =1, A = a] - \psi(a)\Big\} \nonumber \\
      &\quad+ \dfrac{I(S \in \mathcal{S}^k, A = a) \Pr_{k}[S = 0 | X^{(k)}]}{\Pr_{k}[S \in \mathcal{S}^k|X^{(k)}] \Pr_{k}[A = a| X^{k}, I(S \in \mathcal{S}^k) = 1] } \\
      & \quad \quad \quad \times \Big\{ Y - \E_{k}[Y | X^{(k)}, I(S \in \mathcal{S}^k) = 1, A = a]\Big\} \Bigg\} \Bigg)^2 \Bigg]
\end{align*}
and
\begin{align*}
C_{kj} &= \dfrac{1}{\Pr_{k}[S = 0] \Pr_{j}[S = 0]} \E\big[ (\E[Y | X^{(k)}, I(S \in \mathcal{S}^k) =1, A = a] - \psi(a))\\
      &\quad \quad \quad \quad   \times  I(S=0)(\E[Y | X^{(j)}, I(S \in \mathcal{S}^j) =1, A = a] - \psi(a) \big].
\end{align*}
Using this notation rewrite
\[
E[IF^2;w_1, \ldots, w_k] = \sum_{k=1}^K w_k^2 V_k
      + 2 \sum_{k=1}^K \sum_{j=1}^{k-1} w_k w_j  C_{kj}.
\]

To find the optimal weights, defined in terms of minimizing the asymptotic variance, we need to minimize $\E[IF^2;w_1, \ldots, w_k]$ as a function of $(w_1, \ldots, w_k)$ subject to the constraint $\sum_{k=1}^K w_k = 1$. Using a Lagrange multiplier $\lambda$ we rewrite the optimization problem and minimize
\begin{equation}
\label{opt-prob}
T(w_1, \ldots, w_k, \lambda) = \E[IF^2;w_1, \ldots, w_k] - \lambda \left(\sum_{k=1}^K w_k - 1 \right).
\end{equation}
Or equivalently as
\[
T(w_1, \ldots, w_k, \lambda) = \sum_{k=1}^K w_k^2 V_k + 2 \sum_{k=1}^K \sum_{j < k} w_k w_k C_{kj} - \lambda \left(\sum_{k=1}^K w_k - 1 \right).
\]
Both $V_k$ and $C_{kj}$ depend on unknown quantities that can be estimated using plug-in estimators. Denote the estimators by $\widehat V_k$ and $\widehat C_{kj}$. The empirical version of the Lagrance multiplier equation is given by
\begin{equation*}
\widehat T(w_1, \ldots, w_k, \lambda) = \sum_{k=1}^K w_k^2 \widehat V_k + 2 \sum_{k=1}^K \sum_{j < k} w_k w_j \widehat C_{kj} - \lambda \left(\sum_{k=1}^K w_k - 1 \right).
\end{equation*}
Differentiating $\widehat T(w_1, \ldots, w_k, \lambda)$ w.r.t. $w_j$, for $j \in \{1, \ldots, K\}$, gives
\begin{equation}
\label{diff-a}
\frac{\partial \widehat T(w_1, \ldots, w_k, \lambda)}{\partial w_j} = 2 w_j \widehat V_j +2 \sum_{k \neq j} w_k \widehat C_{kj} - \lambda.
\end{equation}
Differentiating $\widehat T(w_1, \ldots, w_k, \lambda)$ w.r.t. $\lambda$ gives 
\begin{equation}
\label{lam}
-\left(\sum_{k=1}^K w_k - 1 \right).
\end{equation}
The optimal weights are obtained by finding the zero crossing of expressions \eqref{diff-a} and \eqref{lam}. 

\clearpage
\section{Sampling model}
\label{app:sep}

Let $D$ be the indicator of whether an individual contributes data to the analyses. Let $s^k_1, \ldots, s^k_{M_k}$ be the trials in $\mathcal{S}^{k}$. By the assumptions made
\[
\E[Y|X^{(k)}, S= s^k_1, A=a] = \ldots = \E[Y|X^{(k)}, S= s^k_{M_k}, A=a] = \E[Y|X^{(k)}, S \in \mathcal{S}^k, A=a].
\]
By the random sampling from the population underlying each trial and from the target population we have
\begin{align*}
\psi_k(a) &= \E_k[\E_k[Y|X^{(k)}, A=a, S \in \mathcal{S}^k]|S=0]\\
&= \E_k[\E_k[Y|X^{(k)}, A=a, S \in \mathcal{S}^k, D = 1]|S=0, D=1].
\end{align*}
Working with the above expression gives
\begin{align*}
\psi_k(a) &= \E_k[\E_k[Y|X^{(k)}, A=a, S \in \mathcal{S}^k, D = 1]|S=0, D=1] \\
&= \E_k\left[\E_k\left[\frac{Y I(A=a, S \in \mathcal{S}^k)}{\Pr[A=a|X^{(k)}, S \in \mathcal{S}^k, D=1] \Pr[S \in \mathcal{S}^k|X^{(k)}, D=1]} \Bigg|X^{(k)}, D = 1\right]\Bigg|S=0, D=1\right] \\ 
&= \frac{1}{\Pr[S=0|D=1]}\E_k\left[\E_k\left[\frac{Y (1 - \Pr[S \in \mathcal{S}^k|X^{(k)}, D=1]) I(A=a, S \in \mathcal{S}^k)}{\Pr[A=a|X^{(k)}, S \in \mathcal{S}^k, D=1] \Pr[S \in \mathcal{S}^k|X^{(k)}, D=1]} \Bigg|X^{(k)}, D = 1\right]\Bigg| D=1\right] \\
&= \frac{1}{\Pr[S=0|D=1]}\E_k\left[\frac{Y (1 - \Pr[S \in \mathcal{S}^k|X^{(k)}, D=1]) I(A=a, S \in \mathcal{S}^k)}{\Pr[A=a|X^{(k)}, S \in \mathcal{S}^k, D=1] \Pr[S \in \mathcal{S}^k|X^{(k)}, D=1]} \Bigg| D=1\right]
\end{align*}  
All quantities in the above expressions condition on the available data ($D=1$). Thus, the potential outcome mean $\psi(a)$ are identifiable under the stratified sampling model.

\clearpage
\section{Additional data analysis results}
\label{app:da}

Table \ref{tab:DA-app} shows the distribution of the covariates in the National Lung Screening Trial (NLST), Prostate, Lung, Colorectal, and Ovarian trial (PLCO), and NHANES datasets (weighted for the NHANES data). Table \ref{tab-da1-app} shows the prevalence rate of lung cancer diagnosis within six years from study enrollment for the NLST, PLCO trials by education level.

\begin{table}[h]
\caption{Summary of participant characteristics in the National Lung Screening Trial (NLST), the Prostate, Lung, Colorectal, and Ovarian trial (PLCO), and the NHANES data (using the sampling probabilities).}
\label{tab:DA-app}
\begin{tabular}{|l|l|l|l|}
\hline
Variable                                                                & NLST        & PLCO        & NHANES      \\ \hline
Age                                                                     & 61.3 (5.0)  & 62.4 (5.2)  & 63.1 (5.5)  \\ \hline
BMI                                                                     & 27.9 (5.1)  & 27.6 (4.8)  & 28.6 (5.6)  \\ \hline
Race (White)                                                            & 90.1\%      & 91.0\%      & 84.5\%      \\ \hline
Education (Some college education)                                                        & 55.5\%      & 53.2\%      & 45.7\%      \\ \hline
Education (High school graduate)                                                        & 38.8\%      & 37.5\%      & 29.1\%      \\ \hline
Smoke years                                                              & 39.6 (7.3)  & 36.1 (9.2) & 42.5 (7.4)  \\ \hline
Gender (Male)                                                           & 58.7\%      & 63.8\%      & 63.0\%      \\ \hline
Marital status (Married)                                                & 68.4\%      & 72.8\%      & 64.7\%      \\ \hline
Pack year                                                               & 55.5 (23.8) & 57.0 (26.2) & 60.6 (28.9) \\ \hline
History of diabetes (Yes)                                                               & 9.3\%      & 8.5\%       & 20.3\%      \\ \hline
History of emphysema (Yes)                                                              & 7.3\%       & 6.5\%       & 8.7\%       \\ \hline
\begin{tabular}[c]{@{}l@{}}History of heart disease \\ or heart attack (Yes)\end{tabular} & 12.2\%      & 13.2\%      & 28.0\%      \\ \hline
History of hypertension (Yes)                                                           & 34.5\%      & 35.3\%      & 45.8\%      \\ \hline
History of asthma (Yes)                                                           & 6.2\%      & NA      & 10.7\%      \\ \hline
Lived with a smoker (Yes)                                                           & 87.5\% & NA & 49.3\%\\ \hline 
Cigarettes per day continuous                                                     & 28.4 (11.4) &  &  28.7 (12.7) \\ \hline 
Cigarettes per day categorical (1-10)                                                     &  & 0.03\% &   \\ \hline 
Cigarettes per day categorical (11-20)                                                     &  & 36.5\% &   \\ \hline 
Cigarettes per day categorical (21-30)                                                     &  & 30.7\% &   \\ \hline 
Cigarettes per day categorical (31-40)                                                     &  & 19.4\% &   \\ \hline 
Cigarettes per day categorical (41-60)                                                     &  & 10.9\% &   \\ \hline 
Cigarettes per day categorical (61-80)                                                     &  & 2.1\% &   \\ \hline 
Cigarettes per day categorical ($>80$ )                                                     &  & 0.4\% &   \\ \hline 
\end{tabular}
\caption*{For both trials, the characteristics are restricted to the chest radiography arms. Continuous covariates are summarized as mean (standard deviation) and categorical covariates are summarized as percentage in each category. BMI is body max index; Smoke years is the total number of years the participant smoked cigarettes; Smoke age is the age at smoking onset; Pack years is calculated as (Total number of years Smoked $\times$ Cigarettes Per Day$/20$). NA indicates that the covariate was not collected. Cigarettes per day were collected as a continuous variable in NHANES and NLST and categorical in PLCO. The summaries for the NHANES data are weighted using the NHANES sampling weights.}
\label{tab:DA-app}
\end{table}

\begin{table}[h]
\caption{Prevalence rate of lung cancer diagnosis within six years from study enrollment for the National Lung Screening Trial (NLST) and the Prostate, Lung, Colorectal, and Ovarian trial (PLCO) trials by education level.}
\label{tab-da1-app}
\begin{tabular}{|l|l|l|}
\hline
                           & NLST  & PLCO  \\ \hline
Did not finish high school & 0.049 & 0.010 \\ \hline
High school degree         & 0.036 &  0.079 \\ \hline
Some college education     & 0.027 & 0.065 \\ \hline
\end{tabular}
\end{table}

\end{document}